\documentclass[12pt]{JHEP3}
\usepackage{amsmath,amssymb,epsfig,bm,amsfonts,yfonts}
\usepackage{units}
\usepackage{cite}

\newcommand{\be}{\begin{equation}}
\newcommand{\ee}{\end{equation}}
\newcommand{\beqa}{\begin{eqnarray}}
\newcommand{\eeqa}{\end{eqnarray}}
 \newcommand{\bn}{\begin{enumerate}}
\newcommand{\en}{\end{enumerate}}

\newcommand{\dr}{\text{d}}

\def\bra#1{\left\langle #1\right|}
\def\eeq{\end{equation}}

\def\ket#1{\left| #1\right\rangle}

\def\Tr{\mathop{\rm Tr}}

\relax

\title{A local Monte Carlo framework for coherent QCD parton energy loss}

\author{
Korinna Christine Zapp${}^{\,1}$, Johanna Stachel${}^{\,2}$ and Urs Achim Wiedemann${}^{\,3}$\\
\vspace{0.1in}

${}^{\,1}$Institute for Particle Physics Phenomenology, Durham University, Durham DH1 3LE, UK\
\vspace{0.1in}

${}^{\,2}$Physikalisches Institut, Universit\"at Heidelberg, Philosophenweg 12, D-69120 Heidelberg, Germany
\vspace{0.1in}

${}^{\,3}$Department of Physics, CERN, Theory Unit, CH-1211 Geneva 23
\vspace{0.1in}

E-mail addresses: {\tt k.c.zapp@durham.ac.uk, stachel@physi.uni-heidelberg.de,Urs.Wiedemann@cern.ch}
}

\abstract{
Monte Carlo (MC) simulations are the standard tool for describing jet-like
multi-particle final states. To apply them to the simulation of medium-modified
jets in heavy ion collisions,  a probabilistic implementation of medium-induced
quantum interference effects is needed. Here, 
we analyze in detail how the quantum interference effects included in 
the Baier-Dokshitzer-Mueller-Peign\'e-Schiff -- Zakharov (BDMPS-Z) formalism of 
medium-induced gluon radiation can be implemented in a quantitatively controlled,
local probabilistic parton cascade.
The resulting MC algorithm is formulated in terms of elastic and inelastic mean free paths,
and it is by construction insensitive to the IR and UV divergences of the total elastic and
inelastic cross sections that serve as its basic building blocks in the incoherent limit. 
Interference effects are implemented by reweighting gluon production histories as a 
function of the number of scattering centers that act within the gluon formation time.
Unlike existing implementations based on gluon formation time, we find generic arguments
for why a quantitative implementation of quantum interference cannot amount to a mere
dead-time requirement for subsequent gluon production. We validate
the proposed MC algorithm by comparing MC simulations with parametric dependencies and 
analytical results of the BDMPS-Z formalism. In particular, we show that the MC algorithm 
interpolates correctly between analytically known limiting cases for totally coherent and 
incoherent gluon production, and that it accounts quantitatively for the medium-induced
gluon energy distribution $\omega \dr I/\dr \omega$ and the resulting average parton energy
loss. We also verify that the MC algorithm implements the transverse momentum broadening
of the BDMPS-Z formalism. We finally discuss why the proposed MC algorithm provides a
suitable starting point for going beyond the approximations of the BDMPS-Z formalism.
}

\preprint{ IPPP/11/16, DCPT/11/32, MCnet-11-11, CERN-PH-TH/2011-043}

\begin{document}
\def\vev#1{\langle#1\rangle}
\def\ov{\over}
\def\le{\left}
\def\ri{\right}
\def\ha{{1\over 2}}
\def\lam{{\lambda}}
\def\Lam{{\Lambda}}
\def\al{{\alpha}}
\def\ket#1{|#1\rangle}
\def\bra#1{\langle#1|}
\def\vev#1{\langle#1\rangle}
\def\det{{\rm det}}
\def\tr{{\rm tr}}
\def\Tr{{\rm Tr}}
\def\NN{{\cal N}}
\def\th{{\theta}}

\def\Om{{\Omega}}
\def \th{{\theta}}

\def \lam {\lambda}
\def \om {\omega}
\def \ra {\rightarrow}
\def \ga {\gamma}
\def\sig{{\sigma}}
\def\ep{{\epsilon}}
\def\apr{{\alpha'}}
\newcommand{\p}{\partial}
\def\LL{{\cal L}}
\def\HH{{\cal H}}
\def\GG{{\cal G}}
\def\TT{{\cal T}}
\def\CC{{\cal C}}
\def\OO{{\cal O}}
\def\PP{{\cal P}}
\def\tir{{\tilde r}}

\newcommand{\bea}{\begin{eqnarray}}
\newcommand{\eea}{\end{eqnarray}}
\newcommand{\nn}{\nonumber\\}

\section{Introduction}
\label{sec1}
%
%
Most generally, the notion 'jet quenching' is currently used to characterize a broad range 
of experimental observations including the modification of high-$p_T$ single inclusive
hadron spectra, jet-like particle correlations and reconstructed jets in nucleus-nucleus
collisions. Jet quenching was discovered at RHIC via measurements of single inclusive
hadron spectra, and the phenomenon was characterized extensively on the level of
two-particle near-side and back-to-back high-$p_T$ correlation functions and particle
yields associated with trigger particles~\cite{Adams:2005dq,Adcox:2004mh}. 
Two-particle correlations displaying very similar features were also seen at the same time
in nucleus-nucleus collisions at the ten times lower center of mass energy of the CERN 
SPS\cite{Agakichiev:2003gg}, 
whereas the single inclusive hadron spectra at the CERN SPS do not show the dramatic 
suppression up to a factor 5 observed at collider energies~\cite{d'Enterria:2004ig}. In recent years,
a strong effort has gone into studying jet quenching at the highest experimentally accessible 
transverse momenta where one may hope to establish the most direct link between the rich
jet quenching phenomenology and a partonic explanation rooted in QCD. 

In this context, we mention that first preliminary results on reconstructed jet measurements have 
become available at RHIC~\cite{Ploskon:2009zd,Bruna:2009em,Lai:2009zq} within the last two years.
With the much wider kinematic reach accessible at the LHC, numerous novel opportunities 
for studying jet quenching emerge now. Data from the first exploratory heavy ion run have 
shown already that the nuclear modification of charged hadron spectra
is somewhat stronger than at RHIC and that it persists beyond $p_T = 20$ GeV~\cite{Aamodt:2010jd}. 
Soon, the kinematic range of these measurements will be extended by a large factor, and
much more detailed information about quenching of high-$p_T$ particles and particle correlations
will become available. Moreover, first measurements of reconstructed jets in heavy ion collisions at LHC 
indicate already that also jets of order $100$ GeV display significant medium-modifications.
In particular, samples of reconstructed dijets display
an energy imbalance distribution that is much wider than in the absence of a nuclear 
environment~\cite{Atlas:2010bu,Chatrchyan:2011sx}. These measurements indicate that the
quenching of reconstructed jets is accompanied by the medium-induced production of many soft 
particles~\cite{CasalderreySolana:2010eh}.
At present, our still incomplete theoretical 
understanding of jet quenching is largely based on the picture that highly
energetic partons produced in dense QCD matter are degraded in energy due to elastic
and inelastic interactions with the surrounding medium prior to hadronization outside the 
medium~\cite{Wiedemann:2009sh,CasalderreySolana:2007zz,Majumder:2010qh,Jacobs:2004qv,Gyulassy:2003mc}.
This picture is supported in particular by data on single inclusive hadron spectra and particle correlations.
The coming years are likely to show a strong interplay of experimental and theoretical efforts to
characterize jet quenching on the level of multi-particle final states and reconstructed jets with the aim of
further constraining the microscopic dynamics of this phenomenon and drawing conclusions about the
properties of the QCD matter by which it is induced. 

Monte Carlo tools have well-recognized advantages for the phenomenological analysis of 
high-$p_T$ multi-particle final states. They are the method of choice for formulating the 
evolution of a parton shower with minimal kinematic approximations and exact implementation 
of conservation laws. They are also best suited for interfacing this partonic evolution with 
the hadronic final state. Moreover, the fact that they generate not only event averages but 
also event distributions of final state
particles meets an obvious experimental demand and allows for the interfacing with modern
jet finding techniques~\cite{Cacciari:2010te}.  To satisfy these experimental and theoretical needs for the study of
heavy ion collisions, several Monte Carlo tools for the simulation of jet quenching
have been developed in recent years. Some of the available tools are full event generators that 
supplement standard 'vacuum' final state parton showers with models of medium-induced gluon radiation 
tailored to analytical calculations of medium-induced 
parton energy loss. \textsc{Hijing}~\cite{Gyulassy:1994ew,Deng:2010mv}, 
\textsc{Q-Pythia}~\cite{Armesto:2009fj}, \textsc{Q-Herwig}~\cite{Armesto:2009ab} and 
\textsc{Pyquen/Hydjet++} \cite{Lokhtin:2005px,Lokhtin:2008xi} fall into this class. Other
approaches modify the \textsc{Pythia} parton shower, e.g., to implement the picture of a 
medium-modified $Q^2$-evolution as in YaJEM~\cite{Renk:2008pp,Renk:2009nz}, 
or to implement rate equations based on a perturbative
calculation of partonic energy loss as in \textsc{Martini}~\cite{Schenke:2009gb}. 
Finally, \textsc{Jewel}~\cite{Zapp:2008gi} aims at formulating a stand-alone final 
state parton shower that interpolates between three analytically known limits, namely the vacuum 
parton shower in the absence of medium effects, the analytically known limit of energy loss
via elastic multiple scattering, and  radiative energy loss. In its current version, however, 
radiative energy loss is modeled similar to other efforts ad hoc in terms of medium-modified splitting 
functions. A more detailed discussion of the current status of MC tools for jet quenching can
be found in Ref.~\cite{Zapp:2010sp}. 

The 'vacuum' parton showers used in MC event generators like \textsc{Pythia}~\cite{Sjostrand:2007gs}, 
\textsc{Herwig}~\cite{Bahr:2008pv} and \textsc{Sherpa}~\cite{Gleisberg:2008ta} are faithful 
representations of the theory of Quantum Chromodynamics (QCD). They resum to leading logarithmic accuracy
the large logarithms associated with collinear gluon emission, and they thus implement with known
accuracy and without additional model-dependent input analytically known features of QCD. 
In contrast, the MC tools for jet quenching listed above are phenomenological models.
They may tailor some numerical steps according to QCD-based analytical calculations, but 
these QCD-based results do not define the MC tool up to controlled accuracy, they solely motivate
physical choices in a more complex (and more complete) dynamical procedure. 
This is a perfectly legitimate approach that meets the demand of a broad range of applications. 
We argue, however, that it is also of interest to complement these pragmatic approaches with a
conceptual exploration of whether a MC algorithm of jet quenching can be formulated as 
a faithful implementation of QCD-based calculations of parton energy loss. Establishing such a clearer
connection between MC tools and analytical QCD-based knowledge of jet quenching may be 
important for constraining the fundamental QCD properties of matter that induce the observed jet quenching 
phenomena. Moreover, as we shall discuss in detail in section~\ref{sec7}, such a faithful MC implementation
provides a suitable starting point for overcoming many of the technical limitations of the state of the
art of analytical parton energy loss calculations. With this motivation, we present in the present
paper a MC tool that provides with controlled accuracy a local and probabilistic implementation of 
the BDMPS-Z formalism of medium-induced radiative parton energy loss. 

The BDMPS-Z formalism~\cite{Baier:1996sk,Baier:1996kr,Zakharov:1997uu,Zakharov:1996fv}
 is historically one of the first QCD-based calculations of medium-induced radiative parton energy 
 loss in the high energy limit. Its path-integral formulation that we recall in section~\ref{sec2}, 
 provides the generating function for formulations of radiative parton energy loss in terms of 
 an opacity expansion~\cite{Wiedemann:2000za,Gyulassy:2000er}. 
 Also, other formulations of medium-induced radiative parton energy loss~\cite{Wang:2001ifa,Arnold:2002ja}
are known to display the same medium-dependencies as the BDMPS-Z formalism
( for a more complete overview, see Ref.~\cite{Wiedemann:2009sh}).
In short, the most widely used radiative parton energy loss calculations are closely related to 
the BDMPS-Z formalism. Moreover, all existing analytical results, as well as generic physics reasoning,
point to the dominant role of the so-called  non-abelian Landau-Pomeranchuk-Migdal (LPM)
effect in medium-induced gluon radiation, and this destructive quantum interference effect
is accounted for in the BDMPS-Z formalism. We therefore expect that a MC implementation of the 
BDMPS-Z formalism can provide more general guidance as to how medium-effects should
be formulated in a MC parton shower. 

We note as an aside, that the BDMPS-Z formalism does not provide
all the information that enters a final-state parton shower. For instance, the BDMPS-Z formalism 
has been derived for a relatively limited kinematic range only (see discussion in section~\ref{sec2}), 
and it does not specify whether and how the angular ordering prescription of a vacuum parton shower 
should be changed in the medium. For recent work on this latter question, see 
Ref.~\cite{MehtarTani:2010ma,MehtarTani:2011tz}. The present paper will not address
these advanced issues. To the extent to which future studies of radiative parton energy loss result in
improvements of the BDMPS-Z formalism, it will be interesting to explore whether these refinements
can be incorporated in modifications of the MC algorithm discussed here. 

A priori, it is unclear whether destructive quantum interference such as the non-abelian LPM effect
can be recast in a  local probabilistic MC implementation of controlled accuracy. A prominent example
in which destructive quantum interference can be formulated indeed in terms of a probabilistic 
prescription is the angular ordering condition of the vacuum parton shower. In general, however,
quantum interference effects need not be in one-to-one correspondence with a local and probabilistic 
procedure. In a previous paper, we had pointed out~\cite{Zapp:2008af} that the concept of 
formation time can be identified unambiguously in the BDMPS-Z formalism and that it could play the same role for the
probabilistic implementation of medium-induced quantum interference as does angular
ordering for implementing destructive interference of gluon production processes in the vacuum. 
In the present paper, this basic idea is worked out in full technical detail.
It will also become clear why some elements of our original proposal have to be modified to arrive at 
a faithful implementation of the BDMPS-Z formalism. 
 
Our paper is organized as follows: We first identify the main building blocks of the proposed 
MC implementation by analyzing in section~\ref{sec2} the BDMPS-Z formalism
in the opacity expansion. Based on this analysis, we discuss in section~\ref{sec3} a simplified MC algorithm 
that does not trace yet the kinematic dependences of parton splitting, but that accounts for the 
formal BDMPS-Z limits of totally coherent and incoherent gluon production on the level of 
total radiated particle yields. Section~\ref{sec4} discusses how this elementary algorithm extends naturally
to a full MC implementation of the BDMPS-Z formalism. In sections ~\ref{sec5} and ~\ref{sec6}, we 
demonstrate that the proposed MC algorithm provides indeed a quantitatively controlled implementation
of the BDMPS-Z formalism. Finally, we discuss in the outlook of section~\ref{sec7} the perspectives for
further uses and developments of this MC tool.

\section{Time-scale for medium-induced interference in the opacity expansion}
\label{sec2}
Medium-induced gluon radiation is expected to be the dominant energy loss
mechanism of
highly energetic partons in QCD matter. Several groups have calculated the 
corresponding medium-induced gluon energy distribution $
\omega\frac{\dr I}{\dr \omega}$ in the kinematical
regime~\cite{Baier:1996sk,Baier:1996kr,Zakharov:1997uu,Zakharov:1996fv,Wiedemann:2000za,Gyulassy:2000er,Wang:2001ifa}

\begin{equation}
	E \gg \omega \gg \vert {\bf k}\vert\, , \vert {\bf q}_i\vert \geq
\Lambda_{\rm QCD}\, ,
	\label{eq2.1}
\end{equation}
where the energy $E$ of the projectile parent parton is much larger than the
energy
$\omega$ of the radiated gluon, which is much larger than its transverse
momentum
${\bf k}$ and the transverse momentum transfers ${\bf q}_i$ from scattering 
centers in the medium. 

 In this section, we recall first that to each order in 
 opacity~\cite{Wiedemann:2000za,Gyulassy:2000er}, the double differential 
 medium-induced gluon distribution $ \omega\frac{\dr I}{\dr \omega\, \dr {\bf k}}$
 can be written in terms of two classes of elementary cross sections (called $R$
and 
 $H$ and defined below), multiplied by weighting factors that
 interpolate between limits of coherent and incoherent particle production. 
 We emphasize that the scales of interpolation between
 coherent and incoherent particle production are set by inverse transverse
energies
 that have an interpretation as formation times. They will play a central role
in the algorithm proposed in section~\ref{sec4}.
 
 \subsection{Medium-induced gluon radiation in the high energy limit}
Our aim is to specify a Monte Carlo algorithm that implements the double
differential 
medium-induced gluon energy distribution  $\omega\frac{\dr I}{\dr \omega\, \dr {\bf k}}$,
derived first by Baier, Dokshitzer, Mueller, Peign\'e and Schiff 
(BDMPS)~\cite{Baier:1996sk,Baier:1996kr} and 
independently by Zakharov~\cite{Zakharov:1997uu,Zakharov:1996fv}
 in the eikonal approximation (\ref{eq2.1}). As a preparatory step, 
we summarize here information about $\omega\frac{\dr I}{\dr \omega\, \dr {\bf k}}$
that will be needed in the following discussion. For a medium of 
finite size, the distribution $\omega\frac{\dr I}{\dr \omega\, \dr {\bf k}}$ of radiated
gluons can be written in the compact path integral 
formulation~\cite{Wiedemann:2000za}
\begin{eqnarray}
  \omega\frac{\dr I}{\dr \omega\, \dr {\bf k}}
  &=& {\alpha_s\,  C_R\over (2\pi)^2\, \omega^2}\,
    2{\rm Re} \int_{\xi_0}^{\infty}\hspace{-0.3cm} \dr y_l
  \int_{y_l}^{\infty} \hspace{-0.3cm} \dr \bar{y}_l\,
   \int \dr {\bf u}\,  
  e^{-i{\bf k}\cdot{\bf u}}   \,
  e^{ -\frac{1}{2} \int_{\bar{y}_l}^{\infty} \dr \xi\, n(\xi)\,
    \sigma({\bf u}) }\,
  \nonumber \\
  && \times {\partial \over \partial {\bf y}}\cdot
  {\partial \over \partial {\bf u}}\,
  \int_{{\bf y}=0}^{{\bf u}={\bf r}(\bar{y}_l)}
  \hspace{-0.5cm} {\cal D}{\bf r}
   \exp\left[ i \int_{y_l}^{\bar{y}_l} \hspace{-0.2cm} \dr \xi
        \frac{\omega}{2} \left(\dot{\bf r}^2
          - \frac{n(\xi) \sigma\left({\bf r}\right)}{i\, \omega} \right)
                      \right]\, .
    \label{eq2.2}
\end{eqnarray}
Here, the right hand side of (\ref{eq2.2}) contains several internal 
variables (${\bf u}$, ${\bf y}$, ${\bf r}$, $y_l$, $\bar{y}_l$), which do not
relate directly to measurable quantities. The longitudinal coordinates $y_l$, $\bar{y}_l$ result
from integrating over the ordered longitudinal gluon emission points in the amplitude
and complex conjugate amplitude of a multiple scattering cross section. The
two-dimensional transverse coordinates ${\bf u}$, ${\bf y}$ and ${\bf r}$ emerge in the
derivation of (\ref{eq2.2}) as distances between the positions of projectile components in
the amplitude and complex conjugate amplitude~\cite{Wiedemann:2000za}.
In the following, we discuss in more detail how the hard 'projectile' parton,
the 'target' medium, and the interaction between both is accounted for by
equation (\ref{eq2.2}).

\underline{Characterization of the medium:}
A partonic projectile that interacts perturbatively with the medium
exchanges gluons with some components of the target. The momentum transfer
between projectile and target can involve both transverse momentum ${\bf q}$
and longitudinal momentum $q_l$. In radiative parton energy loss calculations based
on the high-energy approximation (\ref{eq2.1}), the transverse momentum transfer
dominates, $\vert {\bf q}\vert \gg q_l$. 
This motivates a description of the target in terms
of a collection of
colored static scattering potentials $A({\bf q})$,~\cite{Gyulassy:1993hr,Wang:1994fx} 
\begin{equation}
 \includegraphics[width=12cm]{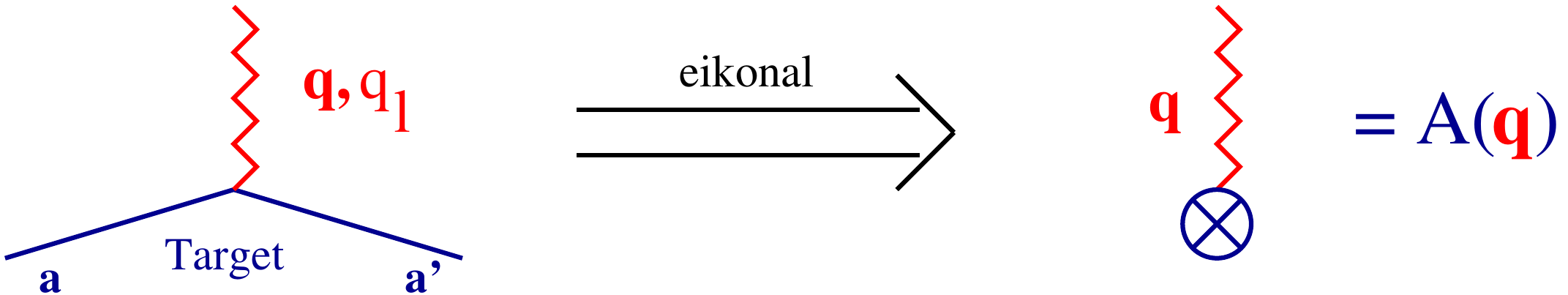}
  \label{eq2.3}
\end{equation}
This approximation neglects recoil effects, and thus it automatically
neglects collisional energy loss. To treat collisional and radiative
energy loss on the same level, one would have to undo the approximation
(\ref{eq2.3}).

In equation (\ref{eq2.2}), the 
scattering potentials $A({\bf q})$ enter the gluon energy distribution 
in the form of the so-called dipole cross section 
\begin{equation}
 \sigma({\bf r}) = 2 \int \frac{\dr {\bf q}}{(2\pi)^2}\, \vert A({\bf q})\vert^2\, 
 	\left(1 - \exp\{ i\, {\bf q}.{\bf r}\} \right)\, .
	\label{eq2.4}
 \end{equation}
 Here, $\vert A({\bf q})\vert^2$ characterizes the differential
 elastic cross section with which the projectile parton transfers a
 transverse momentum ${\bf q}$ to a single scattering center in the medium.
In the gluon energy distribution (\ref{eq2.2}), this quantity is always
multiplied by the density $n(\xi)$ of scattering centers along the
trajectory of the projectile. For notational simplicity, we focus in the following on a 
homogeneous density distribution of scattering centers within a box of length
$L$, that is
\begin{equation}
	n(\xi) =  \Bigg \{ 
	\begin{array}{l}
	n_0\, ,\qquad \hbox{\rm for $0 < \xi < L $\, ,} \\  \\
	0\, ,\qquad \hbox{\rm for $\xi < 0$ or $L< \xi$\, .}
	\end{array} 
	\label{eq2.5}
\end{equation}
Our discussion generalizes to arbitrary density profiles,
but we shall not provide details about this generalization in the present work. 

\underline{Initializing the parent parton:} 
The lower bound $\xi_0$ of the $y_l$-integral of (\ref{eq2.2}) denotes 
the time at which the high energy parton is produced. The parton is
produced either at some finite time, which we set to $\xi_0 = 0$, or it is
produced in the infinite past. These two initializations 
correspond to different physics scenarios:

\begin{itemize}
\item $\xi_0 = 0$\\
If the parton is produced in a hard interaction, then it is produced at a finite
production time, which we set to $\xi_0 = 0$. 
Even in the absence of a medium, partons produced
in a hard collision branch as a consequence of their virtuality. 
Equation (\ref{eq2.2}) contains information about this vacuum splitting,
since it leads in the absence of a medium to
\begin{equation}
	\omega\frac{\dr I(N=0)}{\dr \omega\, \dr {\bf k}} \Big\vert_{\xi_0 = 0}
	= \frac{\alpha_s}{\pi^2} C_R \frac{1}{{\bf k}^2}\, ,
	\label{eq2.6}
\end{equation}
where $C_R = C_F$ for a projectile quark and $C_R = C_A$ for a projectile gluon.
In (\ref{eq2.6}), the notation $N=0$ stands for the zeroth order in the opacity 
expansion, which corresponds to the case $n(\xi) = 0$, where medium effects
vanish.
The result (\ref{eq2.6}) can be identified with the LO $g\to g\, g$ and 
$q\to q\, g$ vacuum splitting functions in the form that these splitting
functions take in the eikonal  limit (\ref{eq2.1}).
\item $\xi_0 = -\infty$\\
The condition $\xi_0 = -\infty$ initializes a parton that  has propagated for
an infinitely long time without branching, prior to possibly interacting with the
medium for
times $\xi \geq 0$. In the absence of a medium, this parton will never branch,
\begin{equation}
	\omega\frac{\dr I(N=0)}{\dr \omega\, \dr {\bf k}} \Big\vert_{\xi_0 = -\infty}=
0\, .
	\label{eq2.7}
\end{equation}
In this sense, the parent parton propagates as if it were 'on-shell'. Because of
confinement,
a colored parton does not propagate forever and this situation will never be 
realized in a physical process in the vacuum. But it is a 
relevant limiting case for understanding the physics contained in (\ref{eq2.2})
\end{itemize}

\underline{Characteristic interaction terms:} 
In the following subsections, we shall demonstrate that the terms related to
{\it vacuum radiation} and {\it medium-induced radiation} can be identified unambiguously
in the radiated gluon energy distribution (\ref{eq2.2}) even outside the incoherent limit.
In preparation for this analysis, we here define the kinematic dependencies
which signal vacuum radiation and medium-induced radiation.
 
Perturbative splittings in the vacuum result in a characteristic 
$\frac{1}{{\bf k}^2}$-distribution of the daughter gluons, with the
transverse momentum measured with respect to the direction of the high energy
parent parton. As {\it vacuum radiation} term, we shall identify the term
\begin{equation}
	H\left({\bf k} \right) = \frac{1}{{\bf k}^2} \, , 
	\label{eq2.9}
\end{equation}
which appears for instance in equation (\ref{eq2.6}). 
Consistent with vacuum radiation, 
this term does not depend on medium properties.
If a gluon, produced by vacuum radiation, scatters incoherently on $N$
scattering 
centers which transfer transverse momenta ${\bf q}_i$ respectively, then the 
transverse momentum distribution of the gluon will be shifted to
\begin{equation}
	H\left({\bf k} +\sum_{i=1}^{N} {\bf q}_i\right) \, .
	\label{eq2.10}
\end{equation}
We will refer also to terms of the form (\ref{eq2.10}) as (shifted) {\it vacuum
radiation}.
 
In the eikonal approximation (\ref{eq2.1}),  the basic cross section for
medium-induced 
gluon radiation in potential scattering with momentum transfer ${\bf q}$ between
target 
and projectile can be written as
\begin{equation}
    \includegraphics[width=12cm]{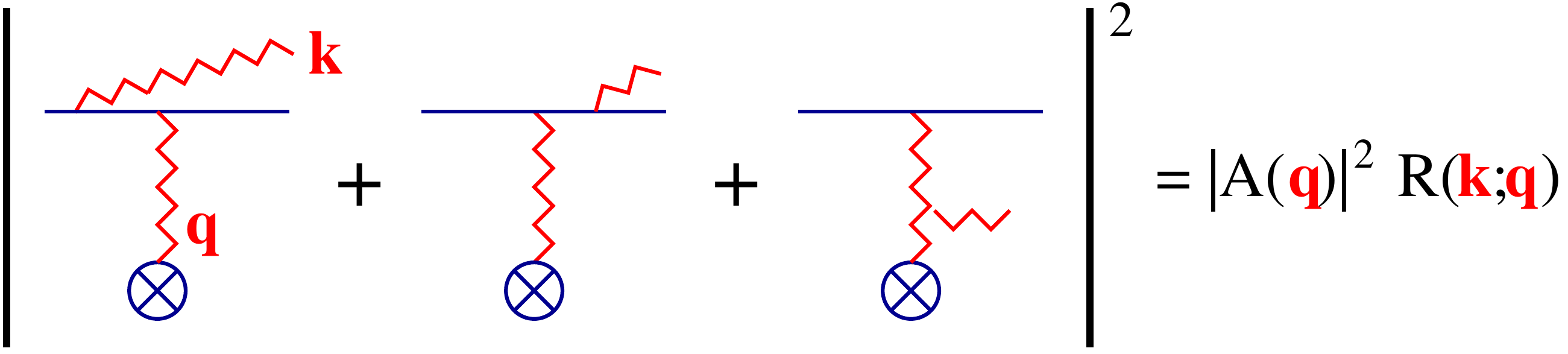}
  \label{eq2.11}
\end{equation}
Here, $\vert A({\bf q})\vert^2$ characterizes the differential elastic
scattering
cross section with which the projectile parton interacts with the static
potential,
and $R({\bf k};{\bf q})$ is the Bertsch-Gunion term~\cite{Gunion:1981qs}
\begin{equation}
	R({\bf k};{\bf q}) = \frac{{\bf q}^2}{{\bf k}^2\left({\bf k}+{\bf q}
\right)^2 }\, ,
	\label{eq2.12}
\end{equation}
which denotes the distribution of gluons of transverse momentum ${\bf k}$,
produced in 
a single incoherent interaction of a high energy parton with a colored
scattering 
potential transferring transverse momentum ${\bf q}$. 
The Bertsch-Gunion term characterizes {\it medium-induced radiation}. 
Consistent with this notation, $R$ vanishes in the absence of
medium effects, that is
for ${\bf q}=0$. If a gluon, after being produced incoherently on one scattering
center, 
scatters subsequently incoherently on $N-1$ other scattering centers, then 
the Bertsch-Gunion momentum distribution is shifted to
\begin{equation}
	R\left({\bf k}+\sum_{i=1}^{N-1} {\bf q}_i;{\bf q}_N\right) \, .
	\label{eq2.13}
\end{equation}
This is the incoherent (i.e. probabilistic) result of multiple elastic
scattering. 

In analyzing the gluon energy distribution (\ref{eq2.2}), we shall also
encounter 
medium-induced radiation terms of the form
\begin{equation}
	R\left({\bf k};\sum_{i=1}^{N} {\bf q}_i \right) \, .
	\label{eq2.14}
\end{equation}
These terms result when $N$ scattering centers act coherently in a single
gluon production process. They will be found in regions of phase space, where the
formation time of the gluon is too long to resolve the $N$ scattering centers
individually. Of course, radiation terms in which gluons are produced in the
coherent scattering on $N$  scattering centers prior to rescattering incoherently
on $M$ other scattering centers can be found also. 
These terms are of the form
\begin{equation}
	R\left({\bf k}+\sum_{j=1}^{M} {\bf q}_{N+j}  ;\sum_{i=1}^{N} {\bf q}_i \right) \, .
	\label{eq2.14b}
\end{equation}

In the following subsections, we analyze the gluon energy distribution
(\ref{eq2.2}) in the opacity expansion. In doing so, we substantiate the
table 
\begin{table}
\begin{center}
\begin{tabular}{|c|c|c|c|}\hline\hline
$\xi_0$  & parent parton &  
$\frac{\dr I^{\rm vacuum}}{\dr \omega \dr {\bf k}}\hfill$ &
$\frac{\dr I^{\rm medium}}{ \dr \omega \dr {\bf k}}\hfill$ interference
of  \\ \hline
 $\xi_0 = -\infty$  & acts as if on-shell &  $= 0$  & terms of form $R$ only \\
$\xi_0= 0 \hfill$  &  branches as if virtual
& $\propto \frac{1}{\omega} \frac{1}{{\bf k}^2}\hfill$ & 
terms of form $R$ and $H$ \\
 \hline\hline
\end{tabular}
 \caption{Summary of characteristics of the $\xi_0=0$ and $\xi_0=-\infty$ cases}
\end{center}
\end{table}
In particular, we specify how interference effects interpolate between 
incoherent elementary processes of the form $H$ and $R$. This will be the
basis for proposing a MC algorithm.  

\subsection{Interference effects for medium-induced gluon radiation (case $\xi_0
= - \infty$)}
\label{sec2.2}

As discussed above, the gluon energy distribution (\ref{eq2.2})
for a parton initialized at time $\xi_0 = - \infty$ 
allows us to study the interference of different sources of medium-induced
radiation in a limiting case, in which complications due to vacuum radiation 
are absent. 

The following analysis relies on the opacity expansion. This is an expansion
of the integrand of (\ref{eq2.2}) in powers of the density of scattering centers
$n(\xi)$
times the effective scattering strength $\sigma({\bf r})$ of a single scattering
center. The opacity expansion 
amounts to an expansion in 
powers of $\int \dr\xi\, n(\xi)\, V_{\rm tot} = n_0\, L\, V_{\rm tot}$, where 
 $V_{\rm tot}$ characterizes the cross section 
presented by the scattering potential $A({\bf q})$,
\begin{equation}
	V_{\rm tot} \equiv  \int \frac{\dr {\bf q}}{(2\pi)^2} \, \vert A({\bf q})
\vert^2\, .
	\label{eq2.15}
\end{equation}
In practice, the opacity expansion of (\ref{eq2.2}) results in integrations
over the transverse momenta ${\bf q}_1$, ...,${\bf q}_N$, which are weighted
by the differential elastic scattering cross sections $\vert A({\bf
q}_1)\vert^2$, ..., 
$\vert A({\bf q}_N)\vert^2$, but which do automatically factorize into powers
of $V_{\rm tot}$. For this reason, the $N$-th order of opacity is obtained
most easily by collecting all terms of order $(n_0\, L)^N$. 
%
\subsubsection{$N=1$ opacity expansion}
\label{sec2.2.1}
The zeroth order in opacity corresponds to the absence of medium effects,
$n(\xi) = 0$, 
when no gluons are radiated,  see equation (\ref{eq2.7}).
The first non-vanishing term in an opacity expansion of (\ref{eq2.2}) 
is then the first order
\begin{equation}
	\omega \frac{\dr I(N=1)}{\dr \omega\, \dr {\bf k}\, \dr {\bf q}_1} =
{\alpha_s\over \pi^2}\, 
   C_R\, \left(n_0\, L\right)
    \frac{1}{(2\pi)^2}\,
    \left( \vert A({\bf q}_1)\vert^2 - V_{\rm tot}\, \bar \delta({\bf q}_1)\,
\right)
    \frac{{\bf q}_1^2}{{\bf k}^2\left({\bf k}+{\bf q}_1 \right)^2 }\, .
  \label{eq2.16}
\end{equation} 
We define $\bar \delta ({\bf q})$ as
\begin{equation}
  \bar \delta ({\bf q}) \equiv (2\pi)^2 \delta({\bf q}) \,,
\end{equation} 
to absorb a factor $(2\pi)^2$ that is common to many formulas in the following. 

In general, to any order in the opacity expansion of (\ref{eq2.2}), factors 
$\vert A({\bf q})\vert^2$ in the integrand appear always in the combination
$\left( \vert A({\bf q})\vert^2 - V_{\rm tot}\, \bar \delta({\bf q}) \right)$.
The terms $V_{\rm tot}$ arise as a consequence of probability conservation,
as we explain in more detail below.  

To first order in opacity, see (\ref{eq2.16}), the term proportional to 
$V_{\rm tot}\, \bar \delta({\bf q}_1)$ vanishes, and the result is of the form
(\ref{eq2.11}) 
of an elastic cross section $\vert A({\bf q})\vert^2$ times a
 Bertsch-Gunion term (\ref{eq2.12}). Hence, the $N=1$ opacity contribution
 to the gluon energy distribution (\ref{eq2.2}) accounts for all radiated
 gluons, which have interacted with exactly one scattering center in the
medium. 
 The prefactor $(n_0\, L)$ in (\ref{eq2.16}) counts
the number of independent gluon productions which occur within the 
length $L$.
%
\subsubsection{$N=2$ opacity expansion and formation time}
\label{sec2.2.2}
Medium-induced quantum interference arises, if a single gluon is produced in
interactions
with at least two scattering centers. In the opacity expansion, this is
realized 
for $N \geq 2$. 
In particular, for $N = 2$, the medium-induced gluon distribution can be written
in the form
\begin{eqnarray}
	\omega \frac{\dr I(N=2)}{\dr \omega\, \dr {\bf k}\, \dr {\bf q}_1\, \dr {\bf q}_2} &=&
{\alpha_s\over \pi^2}\, 
   \frac{C_R}{{(2\pi)^4}}\, 
    \left(\vert A({\bf q}_1)\vert^2 - V_{\rm tot} \bar \delta({\bf
q}_1)\right)\, 
    \left(\vert A({\bf q}_2)\vert^2 - V_{\rm tot} \bar \delta({\bf
q}_2)\right)\, 
    \nonumber \\
    &&\times \left[   
    \frac{\left( n_0\, L\right)^2}{2}\, R({\bf k}+{\bf q}_1; {\bf q}_2)
    	- n_0^2 \frac{1-\cos (LQ_1)}{Q_1^2}  \, R({\bf k}+{\bf q}_1; {\bf q}_2)
	\right.
	\nonumber \\
	&& \qquad \qquad \left.
	+ n_0^2 \frac{1-\cos (LQ_1)}{Q_1^2}  \, R({\bf k}; {\bf q}_1+{\bf q}_2)
	\right]
   \, .
  \label{eq2.17}
\end{eqnarray} 
Here, we have adopted the following conventions~\cite{Wiedemann:2000za}:
To $N$-th order in opacity, subscripts are labeled such that 
$i=1$ is the last, $i = 2$ the next to last and $i=N$ the first scattering
center
along the trajectory of the partonic projectile. Also, the sign of the
transverse 
momenta ${\bf q}_i$ are chosen such that they are flowing from the projectile
to the medium.

The qualitatively novel feature of the $N=2$ result (\ref{eq2.17}), compared to
the
first order result (\ref{eq2.15}), is the appearance of an interference factor
\begin{equation}
	{\cal Z}(Q_1,L) = n_0^2 \frac{1-\cos (LQ_1)}{Q_1^2}\, .
	\label{eq2.18}
\end{equation}
In general, interference factors depend on the in-medium path length $L$ and 
on transverse energies 
\begin{equation}
	Q \equiv \frac{{\bf k}^2}{2\, \omega}\, ,
		\qquad 
		Q_i \equiv \frac{\left( {\bf k}+ \sum_{j=1}^i{\bf q}_j
\right)^2}{2\, \omega}\, .
		\label{eq2.19}
\end{equation}
For the following, it will be useful to view 
the inverse of these transverse energies as formation times. In particular,
\begin{equation}
	\tau  = 1/Q\, ,\qquad \hbox{\rm formation time of the final state
gluon,}
	\label{eq2.20}
\end{equation}
and 
\begin{equation}
	\tau_1  = 1/Q_1\, ,\qquad \hbox{\rm gluon formation time prior to last
interaction at $i=1$\, .}
	\label{eq2.21}
\end{equation}
The interference factor (\ref{eq2.18}) interpolates between the two limiting
cases
\begin{equation}
	n_0^2 \frac{1-\cos (LQ_1)}{Q_1^2}
	= \Bigg \{ 
	\begin{array}{l}
	\frac{(n_0\, L)^2}{2}\, ,\qquad \hbox{\rm for $L \ll \tau_1$\, , $n_0\,
L = {\rm const.}$} \\  \\
	0\, ,\qquad \hbox{\rm for $L \gg \tau_1$\, , $n_0\, L = {\rm const.}$}
	\end{array} 
	\label{eq2.22}
\end{equation}
In both limiting cases, the energy distribution (\ref{eq2.17})  has a
probabilistic 
interpretation:
\begin{itemize}
	\item \underline{Incoherent production limit $L \gg \tau_1$, $n_0\, L =
{\rm const.}$}\\
\begin{eqnarray}
	\omega \frac{\dr I(N=2)}{\dr\omega\, \dr{\bf k}\, \dr{\bf q}_1\, \dr{\bf q}_2 } &=&
{\alpha_s\over \pi^2}\, 
   \frac{C_R}{(2\pi)^4}\,
    \left(\vert A({\bf q}_1)\vert^2 - V_{\rm tot} \bar \delta({\bf
q}_1)\right)\, 
    \left(\vert A({\bf q}_2)\vert^2 - V_{\rm tot} \bar \delta({\bf
q}_2)\right)\, 
    \nonumber \\
    &&\times    \frac{\left( n_0\, L\right)^2}{2}\, R({\bf k}+{\bf q}_1; {\bf
q}_2)
       \, .
  \label{eq2.23}
\end{eqnarray} 
Here the Bertsch-Gunion term $R({\bf k}+{\bf q}_1; {\bf q}_2)$ denotes a
medium-induced 
radiation term, for which the gluon was produced incoherently on the first
scattering center 
with momentum transfer ${\bf q}_2$ and scattered incoherently on the last
scattering center with momentum transfer ${\bf q}_1$ . 
\item \underline{Totally coherent production limit $L \ll \tau_1$, $n_0\, L =
{\rm const.}$}\\
\begin{eqnarray}
	\omega \frac{\dr I(N=2)}{\dr\omega\, \dr{\bf k}\, \dr{\bf q}_1\, \dr{\bf q}_2 } &=&
{\alpha_s\over \pi^2}\, 
   \frac{C_R}{(2\pi)^4}\,
    \left(\vert A({\bf q}_1)\vert^2 - V_{\rm tot} \bar \delta({\bf
q}_1)\right)\, 
    \left(\vert A({\bf q}_2)\vert^2 - V_{\rm tot} \bar \delta({\bf
q}_2)\right)\, 
    \nonumber \\
    &&\times 
    \frac{\left( n_0\, L\right)^2}{2}\,  R({\bf k}; {\bf q}_1+{\bf q}_2)
	   \, .
  \label{eq2.24}
\end{eqnarray} 
Here, the Bertsch-Gunion term 
$R({\bf k}; {\bf q}_1+{\bf q}_2)$ denotes a coherent gluon production 
in which the two scattering centers are not resolved but act effectively
as a single one. 
\end{itemize}
In the expressions above, there are terms proportional to $\vert A({\bf
q}_1)\vert^2\, \vert A({\bf q}_2)\vert^2$. These correspond to processes, in
which the radiated gluon 
exchanges momentum with exactly two scattering centers. 
In addition, there are terms proportional to  $V_{\rm tot}\, \vert A({\bf
q}_2)\vert^2$,
in equations (\ref{eq2.23}) and (\ref{eq2.24}), 
which involve only one momentum transfer with the target. 
For these latter terms, the totally coherent and incoherent limits differ by a
factor 2. 
This can be understood in terms of a probabilistic picture of the partonic
dynamics: 
In the incoherent case,  the gluon can scatter on the second scattering center 
at $\xi_1$ only after it was produced incoherently at position $\xi_2$. The
corresponding weight from the integrals along the trajectory is 
$\propto \int_0^L \dr\xi_2\, n_0 \int_{\xi_2}^L \dr\xi_1\, n_0 = (n_0\, L)^2/2$. 
In contrast, in the coherent case when both scattering centers lie within the 
formation time of the gluon, their time
ordering does not matter and the probability conserving contribution has the 
weight $\int_0^L \dr\xi_2\, n_0 \int_0^L \dr\xi_1\, n_0 = (n_0\, L)^2$,
which is a factor 2 larger.

\subsection{Combining medium-induced and vacuum gluon radiation 
(case $\xi_0 = 0$)}
\label{sec2.3}

In section~\ref{sec2.2}, we discussed how destructive interference gives
rise to formation time scales in the gluon energy distribution (\ref{eq2.2}) 
with initialization $\xi_0 = -\infty$, where vacuum radiation is absent. Here, 
we parallel this discussion for the initialization $\xi_0 = 0$, when the hard 
projectile splits also in the absence of a medium, as expected for a virtual
state.

\subsubsection{$N=1$ Opacity expansion for $\xi_0 = 0$}
To zeroth order in opacity, the gluon energy distribution (\ref{eq2.2}) yields
the singular part $\frac{\dr I}{{\dr z\, \dr{\bf k}}} = \alpha_s C_R \frac{1}{{\bf k}^2} \frac{1}{z}$.
This is the leading order  quark
or gluon splitting 
function for $z= \omega/E$ within the eikonal approximation (\ref{eq2.1}).

Destructive interference arises already to first order in opacity, 
\begin{eqnarray}
  \omega \frac{\dr I(N=1)}{ \dr \omega\, \dr {\bf k}\, \dr {\bf q}_1}
  &=& {\alpha_s\over \pi^2}\, \frac{C_R}{(2\pi)^2}\, 
    \left( |A({\bf q}_1)|^2 - V_{\rm tot}\, \bar \delta({\bf q}_1) \right)
    \, n_0\frac{L\, Q_1 - \sin\left(L\, Q_1 \right)}{Q_1}
    	\nonumber \\
	&& \qquad \qquad \times
    \, \left[  \frac{1}{\left({\bf k}+{\bf q}_1\right)^2} 
    	+ \frac{{\bf q}_1^2}{{\bf k}^2\, \left({\bf k}+{\bf q}_1\right)^2} 
\right]\, .
  \label{eq2.31}
\end{eqnarray} 
Here, $\tau_1 = 1/Q_1$ is the formation time of the gluon prior to scattering
on the medium with momentum transfer ${\bf q}_1$. The limiting cases are:

\begin{itemize}
\item \underline{The limit $L \ll \tau_1$, $n_0\, L = {\rm const.}$}\\
One finds the limit
\begin{eqnarray}
	\lim_{L/\tau_1\to  0}   \omega \frac{\dr I(N=1)}{\dr
\omega\, \dr {\bf k}\, \dr {\bf q}_1 } = 0\, .
	\label{eq2.32}
\end{eqnarray} 
As we discuss in more detail in appendix~\ref{appa}, 
this limit is consistent with the probabilistic picture that a gluon can only be 
produced in a scattering if it is formed as part of the incoming projectile wave function
prior to the scattering. 
	\item \underline{$N=1$: Incoherent production limit for $L \gg \tau_1$,
$n_0\, L = {\rm const.}$}\\
\begin{eqnarray}
	\omega \frac{\dr I(N=1)}{\dr \omega\, \dr{\bf k}\, \dr{\bf q}_1 } &=&
{\alpha_s\over \pi^2}\, 
   \frac{C_R}{(2\pi)^2}\, \left(n_0\, L \right)
   \left[ - V_{\rm tot} \bar \delta({\bf q}_1) H({\bf k})
   + \vert A({\bf q}_1)\vert^2  \, H({\bf k}+{\bf q}_1)
   \right.
   \nonumber \\
   && \qquad \qquad \qquad \qquad \left.    
   + \vert A({\bf q}_1)\vert^2 \,  R({\bf k},\, {\bf q}_1)
    \right]  
    \label{eq2.33}
\end{eqnarray} 
On the right hand side of this equation, the first term is proportional to
$V_{\rm tot}$ and 
implements probability conservation: 
the total probability that a scattering with some momentum transfer ${\bf q}_1$
occurs
is subtracted from the $N=0$-contribution that no momentum transfer occurs. 
If a momentum transfer occurs, then this momentum transfer can either shift
probabilistically
the transverse momentum of the fully formed gluon. This is the second term
proportional
to $H({\bf k} + {\bf q}_1)$. Alternatively, the momentum transfer leads to a
medium-induced
gluon production, distributed according to the Bertsch-Gunion term $R({\bf k},\,
{\bf q}_1)$.
\end{itemize}
%

\subsubsection{$N=2$ Opacity expansion for $\xi_0 = 0$}
\label{sec2.3.2}

For $\xi=0$,
the 2nd order in opacity of equation (\ref{eq2.2}) can be written in the compact form~\cite{Wiedemann:2000za}
\begin{eqnarray}
	\omega \frac{\dr I(N=2)}{\dr\omega\, \dr{\bf k}\, \dr{\bf q}_1\, d{\bf q}_2} &=& 
	{\alpha_s\over \pi^2}\, 
   \frac{C_R}{(2\pi)^4}     \left( |A({\bf q}_1)|^2 - V_{\rm tot}\, \bar 
\delta({\bf q}_1) \right)
       \left( |A({\bf q}_2)|^2 - V_{\rm tot}\, \bar \delta({\bf q}_2) \right)
\nonumber\\
       && \times  \frac{1}{\left(2\, \omega\right)^2} \sum_{j=0}^2 \, {\cal Z}_{j+1}
       \left({\bf k}+\sum_{i=1}^2 {\bf q}_i \right)\cdot\left({\bf k}+\sum_{i=1}^{2-j} {\bf q}_i \right)\, ,
       \label{eq2a}
\end{eqnarray}
where
\begin{eqnarray}
	{\cal Z}_1 &=& n_0^2 \frac{2\cos(LQ_2) - 2 + L^2Q_2^2}{2\, Q_2^4} \, , \\
	{\cal Z}_2 &=& n_0^2 \frac{Q_1^3 \left[2\cos(LQ_2) - 2 + L^2Q_2^2\right]}{
					2\, Q_1^3  \left(Q_1-Q_2\right) Q_2^3}
					- n_0^2 \frac{Q_1^3 \left[2\cos(LQ_1) - 1 + L^2Q_1^2\right]}{
					2\, Q_1^3  \left(Q_1-Q_2\right) Q_2^3}\, ,
					\label{eqz2} \\
	{\cal Z}_3 &=& n_0^2 \frac{Q_1^2\left[ -1 + 2\cos(LQ_2)\right] }{
					Q Q_1^2 \left(Q_1-Q_2\right) Q_2^2}
						- n_0^2 \frac{Q_2^2\left[ -1 + 2\cos(LQ_1)\right] }{
					Q Q_1^2 \left(Q_1-Q_2\right) Q_2^2}\, .
\end{eqnarray}
We consider again the case of a fixed number of effective scattering centers, $n_0\, L = {\rm const}$.
In the limit $n_0\, L = {\rm const}$, $L\to 0$, expression (\ref{eq2a}) vanishes, 
\begin{eqnarray}
	\lim_{n_0\, L = {\rm const}\, ,\, L\to 0}\, 
	\omega \frac{\dr I(N=2)}{\dr\omega\, \dr{\bf k}\, \dr{\bf q}_1\, \dr{\bf q}_2 } 
	= 0\, ,
	 \label{eq2.34a}
\end{eqnarray}
and so do all higher orders in opacity. 
In the opposite limit, $n_0\, L = {\rm const}$, $L\to \infty$, one finds the totally incoherent limit 
\begin{eqnarray}
	\lim_{n_0\, L = {\rm const}\, ,\, L\to \infty}\,
	\omega \frac{\dr I(N=2)}{\dr\omega\, \dr{\bf k}\, \dr{\bf q}_1\, \dr{\bf q}_2 } 
      &=&  {\alpha_s\over \pi^2}\, 
   \frac{C_R}{(2\pi)^4}
   \nonumber \\
   &&  \left[ 
   \frac{\left(n_0\, L \right)^2}{2} \vert A({\bf q}_1)\vert^2\, \vert A({\bf
q}_2)\vert^2\,  \left[ 
   H({\bf k} + {\bf q}_1+{\bf q}_2) + 
    R({\bf k}+{\bf q}_1,{\bf q}_2) \right] \right. \nonumber \\
    && - V_{\rm tot} \bar \delta({\bf q}_1)  \frac{\left(n_0\, L \right)^2}{2}
   \vert A({\bf q}_2)\vert^2
   \left[ H({\bf k} + {\bf q}_2) + R({\bf k};{\bf q}_2)\right]
    \nonumber \\
   && \left. + \frac{\left(n_0\, L\, V_{\rm tot} \right)^2}{2} \bar \delta({\bf
q}_2)
\bar \delta({\bf q}_1) H({\bf k})
   \right]\, .
    \label{eq2.34}
\end{eqnarray} 
The probabilistic interpretation of this expression is as follows:
If the gluon has interacted incoherently with two scattering centers prior to
escaping from the medium after length $L$ with momentum ${\bf k}$, then 
this gluon was either produced in a vacuum splitting and accumulated 
transverse momentum incoherently in two scattering. 
This is the term $H({\bf k} + {\bf q}_1+{\bf q}_2)$. Alternatively, the gluon
was produced in a medium-induced interaction $R({\bf k}+{\bf q}_1,{\bf q}_2)$
with momentum transfer ${\bf q}_2$ and accumulated additional
transverse momentum ${\bf q}_1$ incoherently in a second interaction. 
The second and third line of (\ref{eq2.34}) readjust the probabilities that the gluon was
produced with less than two momentum transfers from the medium. In particular, to all orders
in $N$, the vacuum emission $H({\bf k})$ remains unmodified by the medium with the
weight given by the no-scattering probability $S= \exp\left[ - n_0\, L\, V_{\rm tot}
\right]$, and the last line is the second order in opacity of $S\,  H({\bf k})$. Similarly, the
second line readjusts
the probability for gluon production processes with exactly one scattering
center involved. 

What dictates the scale at which the vanishing (totally coherent) radiation pattern (\ref{eq2.34a}) 
evolves into a fully  developed  incoherent radiation pattern (\ref{eq2.34})? For reasons that will 
become clear in the following subsection, we focus our discussion of this question on the 
medium-induced radiation term  $R({\bf k}+{\bf q}_1,{\bf q}_2)$. We observe that in the limit 
$n_0\, L = {\rm const}$, $L\to \infty$ of the gluon energy distribution (\ref{eq2a}), only the term
proportional to ${\cal Z}_2$ contributes to the medium-induced radiation term $R$. 
The limiting cases of ${\cal Z}_2$ are 
\begin{equation}
	\lim_{n_0\, L = {\rm const}\, ,\, L\to \infty}\, 
	{\cal Z}_2 
	 =  \frac{n_0^2L^2}{2Q_1Q_2}\, ,
	\qquad
	\lim_{n_0\, L = {\rm const}\, ,\, L\to 0}\, 
	{\cal Z}_2 
	 =  0\, .
\end{equation}
Inspection of equation (\ref{eqz2}) shows that for $n_0\, L = {\rm const}$, 
the first term vanishes for scales $L \ll 1/Q_2$ and the second term for length
scales $L \ll 1/Q_1$. To fully explore the physical implications of this observation,
we recall that $Q_2$ is the transverse energy of the gluon prior to interacting with
the target, and $Q_1$ is the transverse energy of the gluon after the first and prior to
the second scattering. For the most likely scattering histories, transverse energy will be built up 
step by step in multiple scattering,  $Q_2 \ll Q_1$. We have written this as a strong inequality
with the idea that medium-induced transverse momentum broadening
should dominate over the initial transverse momentum of the vacuum radiation. 
Now, for $Q_2 \ll Q_1$, one sees that the second term in (\ref{eqz2}) dominates the
value of ${\cal Z}_2$ for sufficiently large $L$, and this second term dies out on length
scales $L \ll 1/Q_1$. This leads us to the qualitative conclusion that it is the formation time
$1/Q_1$ of the gluon prior to its last interaction with the target that determines whether the
radiation $R$ takes place. The gluon is only radiated if its formation time is sufficiently short
so that formation is completed on a scale comparable with the in-medium path length. 

\subsubsection{Guidance for an MC implementation}
\label{sec2.3.3}

A remarkable simplification of MC simulations of  the ${\bf k}$-integrated radiation 
pattern arises from the fact that vacuum terms like $H({\bf k}+{\bf q})$ in (\ref{eq2.33}) do
not contribute to parton energy loss. This is so, since $H({\bf k}+{\bf q})$ amounts to a 
probability-conserving redistribution of gluons in transverse momentum space; this 
redistribution affects neither the yield of emitted gluons, nor their energy distribution. As
a consequence, neglecting the terms proportional to $H$ does not affect the gluon
energy distribution $\omega \dr I/\dr\omega$. For ${\bf k}$-differential distributions, a 
similar a priori argument does not exist. We note as an aside that terms proportional to $H$ were not
taken into account in the original derivation of the BDMPS-Z formalism. They
appeared first in the derivation of Ref.~\cite{Wiedemann:2000za} that leads to (\ref{eq2.2}).
That they modify  the transverse momentum distribution was also
recognized in Ref.~\cite{Baier:2001qw}. However, there is numerical evidence
that inclusion of these terms is a numerically small effect~\cite{Wiedemann:2000za}.
Based on this observation, {\it we shall seek a MC implementation of the
BDMPS-Z formalism that neglects terms proportional to $H$}. This treatment is exact
for ${\bf k}$-integrated quantities, and - as we shall show in section~\ref{sec6} - 
it is a satisfactory approximation for ${\bf k}$-differential information.

For the medium-induced radiation terms $R$, at first order in opacity, 
the only difference between the cases  $\xi_0 = -\infty$  (\ref{eq2.16}) and $\xi_0 = 0$  
(\ref{eq2.31}) is the reduction in the phase space of $R$ due to the destructive interference 
term $n_0\left( {L\, Q_1 - \sin \left(L\, Q_1\right)}\right) / Q_1$. The analysis to first order in
opacity did not allow us to disentangle between an interpretation of this phase space cut in
terms of either i) the formation time prior to the very first or ii) prior to the very last interaction 
with the medium. The analysis of the 2nd order in opacity, however, gave support to the
second interpretation, see section~\ref{sec2.3.2}.
Motivated by this observation, {\it we shall propose in section~\ref{sec4} a Monte
Carlo implementation of the BDMPS-Z formalism for $\xi_0 = 0$, according to which gluons
are rejected from the simulation if their formation is not completed within the medium}. 

The analysis of the opacity expansion in section~\ref{sec2.3.2} supports only 
the parametric statement that those medium-induced gluons contribute to the distribution (\ref{eq2.2}) 
whose formation is completed on a length scale comparable to $L$. 
It is one conceivable (though not unique) implementation of this parametric argument 
to count solely gluons whose formation is 
completed within the medium. We note that in establishing
a one-to-one correspondence between the opacity expansion of (\ref{eq2.2}) and a MC algorithm,
this is the only point where we have found only parametric and not quantitative guidance.
Accordingly, we have tested numerically some variations of this prescription,
and we shall comment on this in section~\ref{sec5}.

\section{ A simplified problem: a MC algorithm for $\langle N_g\rangle$ 
in the totally coherent and incoherent BDMPS limits } 
\label{sec3}
The main aim of this paper is to formulate a MC algorithm that interpolates 
correctly between the analytically known BDMPS results in the opacity expansion.
Explicit expressions for these limits are known analytically~\cite{Wiedemann:2000za}
to  arbitrary high orders in opacity. For the case of an incident projectile 
($\xi_0=-\infty$), the totally coherent limit is
\begin{equation}
	\omega \frac{\dr I}{\dr\omega\, \dr{\bf k}} \Bigg\vert^\text{coh}
	=  {\alpha_s\over \pi^2}\, C_R\, \exp\left[-n_0\, L\, V_{\rm tot} \right]\, 
		\sum_{N_s=1}^\infty \frac{1}{N_s!}
		\left( \prod_{i=1}^{N_s} \int_{q_i} n_0\, L \right)\,
		R\left( {\bf k}, \sum_{j=1}^{N_s} {\bf q}_j \right)\, ,
	\label{eq3.1}
\end{equation}
and the incoherent limit is
\begin{equation}
	\omega \frac{\dr I}{\dr \omega\, \dr{\bf k}} \Bigg\vert^\text{incoh}
	=  {\alpha_s\over \pi^2}\, C_R\, \exp\left[-n_0\, L\, V_{\rm tot} \right]\, 
		\sum_{N_s=1}^\infty \frac{1}{N_s!}
		\left( \prod_{i=1}^{N_s}   \int_{q_i} n_0\, L \right)\,
		\sum_{j=1}^{N_s} R\left( {\bf k} + \sum_{l=1}^{j-1} {\bf q}_l, {\bf q}_j \right)\, .
		\label{eq3.2}
\end{equation}
Here, we have used the shorthand
\begin{equation}
	\int_{q_i} f(q) \equiv \int \frac{\dr{\bf q}_i}{(2\pi)^2} \, \vert A({\bf q}_i)\vert^2 \, f(q)\, .
	\label{eq3.3}
\end{equation}
In general, contributions to $N$-th order in opacity contain products of a number $N_s$ 
($1 \leq N_s \leq N$) of cross sections $\vert A({\bf q}_i)\vert^2$, 
and a number $N-N_s$ of cross sections $V_{\rm tot}$, obtained from expanding the prefactor
$\exp\left[ -n_0\, L\, V_{\rm tot} \right]$. 

In this section, we consider first the simpler problem of formulating for the limits of totally
coherent and incoherent gluon production an algorithm for the
momentum-space integrated average number of radiated gluons,
\begin{equation}
	\langle N_g\rangle = \int_{{\bf k}, \omega}\, 
	 \frac{\dr I}{\dr\omega\, \dr{\bf k}} \, .
	 \label{eq3.4}
\end{equation}
This study will be extended to the differential spectrum in section~\ref{sec4}.
%

\subsection{Relating BDMPS-Z to elastic and inelastic mean free paths}
\label{sec3.1}
We consider first the $N_s = 1$ scattering contribution to the totally 
coherent and incoherent BDMPS limits (\ref{eq3.1}) and (\ref{eq3.2}). The
resulting average number of radiated gluons is
\begin{eqnarray}
	\langle N_g\rangle (N_s=1) &=& \exp\left[-n_0\, L\, V_{\rm tot} \right]\, 
	n_0\, L \, {\alpha_s\over \pi^2}\, C_R\, 
	     \int_{{\bf k}, \omega}
		\int  \frac{\dr{\bf q}}{(2\pi)^2}  \vert A({\bf q})\vert^2\, 
		\frac{1}{\omega}
		R\left( {\bf k},  {\bf q} \right)
		\nonumber \\
		&\equiv&
			n_0\, L\, \sigma_{\rm inel}
			 \exp\left[- n_0\, L\, V_{\rm tot}  \right]\, .
			 \label{eq3.5}
\end{eqnarray}
Here, we have used the analysis of equation (\ref{eq2.16}) to define the inelastic cross section 
for incoherent gluon production on a single scattering center as 
\begin{equation}
	\sigma_{\rm inel} \equiv 
	{\alpha_s\over \pi^2}\, C_R\, 
	     \int_{{\bf k}, \omega}
		\int  \frac{\dr{\bf q}}{(2\pi)^2}  \vert A({\bf q})\vert^2\, 
		\frac{1}{\omega}
		R\left( {\bf k},  {\bf q} \right)\, .
\end{equation}
Here, the integrations over ${\bf k}$ and  $\omega$ require regularization. 
The value of the regulator is a physical choice: it determines up to which
 soft scale infrared and collinear production processes are counted towards the inelastic 
 cross section. We shall explain in section~\ref{sec5} how, based on this definition of $\sigma_{\rm inel}$,
 one can calculate measurable quantities that are insensitive to the choice of regulators. 
In the BDMPS-Z formalism, factors $\vert A({\bf q})\vert^2$ and $V_{\rm tot}$  are always multiplied 
by the density $n_0$ of scattering centers.
The product $n_0\, \sigma_{\rm inel}$ defines the inelastic mean free path $\lambda_{\rm inel}$
\begin{equation}
	   \frac{L}{\lambda_{\rm inel}} \equiv n_0\, L\, \sigma_{\rm inel}\, .
	   \label{eq3.6}
\end{equation}
Physical results depend on 
$\lambda_{\rm inel}$, but they do not depend separately on $\sigma_{\rm  inel}$ and $n_0$. 

As seen in the discussion of (\ref{eq2.3}),  the term $\vert A({\bf q})\vert^2$ can be viewed as the differential
elastic cross section $\frac{\dr\sigma_{\rm el}}{\dr{\bf q}}$ for scattering of the partonic projectile on a single
target.
Accordingly, we identify
\begin{equation}
 V_{\rm tot} = \int \frac{\dr{\bf q}}{(2\pi)^2} \, \vert A({\bf q})\vert^2  \equiv \sigma_{\rm el}\, .
    \label{eq3.7}
\end{equation}
The exponential factor $ \exp\left(- n_0\, L\, V_{\rm tot} \right)$ can then be written in terms of the 
elastic mean free path $\lambda_{\rm el}$,
\begin{equation}
	 \exp\left(- n_0\, L\, V_{\rm tot} \right) \equiv  \exp\left(- \frac{L}{\lambda_{\rm el}} \right)\, .
	 \label{eq3.8}
\end{equation}

\subsubsection{Incoherent limit}
\label{sec3.1.1}
The higher order terms of the 
coherent and totally incoherent BDMPS-Z limits (\ref{eq3.1}) and (\ref{eq3.2}) differ. 
In particular, for $N_s=2$, we have
\begin{eqnarray}
	\langle N_g^{\rm incoh}\rangle (N_s=2) &=& 
	 {\alpha_s\over \pi^2}\, C_R\, e^{- n_0\, L\, V_{\rm tot} }\, 
		\frac{1}{2!} \int_{{\bf k}, \omega}
		 \int_{q_1}  \int_{q_2} \left(n_0\, L \right)^2\,
		 \left(  \frac{R\left( {\bf k} , {\bf q}_1 \right)}{\omega} +
		 \frac{R\left( {\bf k} + {\bf q}_1, {\bf q}_2 \right)}{\omega}  \right)
		 \nonumber \\
		 &=& e^{- L/\, \lambda_{\rm el} }\, 
		\frac{1}{2!}  L^2\, \frac{2}{\lambda_{\rm el}\, \lambda_{\rm inel}}\, .
		\label{eq3.9}
\end{eqnarray}
Here, the first term $\propto R\left( {\bf k} , {\bf q}_1 \right)$ has a ${\bf q}_2$-independent
integrand and can be written as a factor $1/\lambda_{\rm el}$. This is a consequence
of $\int_{q_2} 1 = V_{\rm tot}$ and the argument leading
to (\ref{eq3.8}). For the second term $\propto R\left( {\bf k} +{\bf q}_1, {\bf q}_2 \right)$,
a formal shift ${\bf k} \to {\bf k}-{\bf q}_1$ in the integral of (\ref{eq3.9})
indicates that its contribution to the transverse momentum integrated average
(\ref{eq3.9}) is of the same magnitude. This prompts us to identify in the 
incoherent limit the higher orders of $N_s$ with
\begin{eqnarray}
	\langle N_g^{\rm incoh}\rangle (N_s) 
			 &=& e^{- L/ \lambda_{\rm el} }\, 
		\frac{1}{N_s!}  L^{N_s}\, \frac{N_s}{\lambda_{\rm inel}\, \lambda^{N_s-1}_{\rm el}}\, .
		\label{eq3.10}
\end{eqnarray}
Summing over all orders of $N_s$, one finds

\begin{eqnarray}
	\langle N_g^{\rm incoh}\rangle = \sum_{N_s=1}^{\infty}
	\langle N_g^{\rm incoh}\rangle (N_s) 
			 &=& 
		 \frac{L}{\lambda_{\rm inel}}\, .
		\label{eq3.11}
\end{eqnarray}
This is the expected result for the average number of gluons produced incoherently within 
a length $L$, and it thus supports our identification of momentum-integrated terms in the 
BDMPS-Z formalism with elastic and inelastic mean free paths. 

\subsubsection{Totally coherent limit}

To arbitrary order in opacity, we find from (\ref{eq3.1}) for the totally coherent limit
\begin{eqnarray}
	\langle N_g^{\rm coh} \rangle (N_s) &=& e^{- L/\lambda_{\rm tot}} \, 
		\frac{1}{N_s!} 
		\left( \frac{L}{\lambda_{\rm el}} \right)^{N_s} 
		 \int_{{\bf k}, \omega}
		\left( \prod_{i=1}^{N_s} \left( \int  \frac{d{\bf q}}{(2\pi)^2}  
			\frac{\vert A({\bf q})\vert^2}{V_{\rm tot}} \right) \right)\, 
		{\alpha_s\over \pi^2}\, \frac{C_R}{\omega}
		R\left( {\bf k},  \sum_{j=1}^{N_s} {\bf q}_j \right)
		\label{eq3.12}
\end{eqnarray}
In general, the ${\bf k}$-integration over $R\left( {\bf k},  \sum_{j=1}^{N_s} {\bf q}_j \right)$
differs from the integration over $R\left( {\bf k}+  \sum_{l=1}^{j-1} {\bf q}_l, {\bf q}_j \right)$
in the incoherent limit. However, both ${\bf k}$-integrals are dominated by 
contributions from the two (IR regulated) singularities in the Bertsch-Gunion factor,
and these dominant contributions are identical for both integrals. This prompts us
to write
\begin{equation}
	\langle N_g^{\rm coh} \rangle (N_s) = 
	\exp\left(- \frac{L}{\lambda_{\rm el}} \right)\, 
		\frac{1}{N_s!} 
		\left( \frac{L}{\lambda_{\rm el}} \right)^{N_s} 
		\frac{\lambda_{\rm el}}{\lambda_{\rm inel}}\, .
			 \label{eq3.13}
\end{equation}
All totally coherent contributions $\langle N_g^{\rm coh} \rangle (N_s)$ are exactly 
one factor $1/N_s$ smaller than those of the incoherent limit (\ref{eq3.10}).
The resulting average number of gluons produced totally coherently is
\begin{equation}
	\langle N_g^{\rm coh}\rangle = \sum_{N_s=1}^{\infty}
	\langle N_g^{\rm coh}\rangle (N_s) =
		 \frac{\lambda_{\rm el}}{\lambda_{\rm inel}}
		 \left(1 - e^{-L/\lambda_{\rm el}} \right)\, .
		\label{eq3.14}
\end{equation}
%
\subsubsection{Ambiguities in identifying mean free paths in the BDMPS-Z formalism}
\label{sec3.1.3}
In the BDMPS-Z formalism, one calculates radiation cross sections for multiple scattering
processes that have one additional gluon in the final state and that involve a very large number
of elastic interactions. Therefore, the BDMPS-Z formalism is derived under the assumption that
$\lambda_{\rm el} \ll \lambda_{\rm inel}$. The ratio of these mean free paths sets the value
of the strong coupling constant, $\lambda_{\rm el} / \lambda_{\rm inel} \propto
\alpha_s$,
see section~\ref{sec5} for a quantitative discussion.  In this sense, the
BDMPS-Z formalism
is a weak coupling approach with regards to gluon radiation, whereas it resums the possibly
non-perturbatively strong interactions between projectile and target. 

In general, the total mean free path $\lambda_{\rm tot}$ is defined as 
\begin{equation}
	\frac{1}{\lambda_{\rm tot}} = \frac{1}{\lambda_{\rm el}} +  \frac{1}{\lambda_{\rm inel}}\, .
\end{equation}
However, in a formalism where $\lambda_{\rm el} / \lambda_{\rm
inel} = O(\alpha_s) \ll 1$, 
the inverse of $\lambda_{\rm tot}$ equals the inverse of $\lambda_{\rm el}$ up to subleading
corrections of $O(\alpha_s)$ that become negligible. That leads to some ambiguities in
identifying mean free paths in the BDMPS-Z formalism. 
In the discussion so far, we have chosen to interpret $V_{\rm tot}$ as a phase-space integrated
elastic cross section.  This is natural in the light of equation (\ref{eq2.3}). On the other hand, 
one has also the choice of identifying $n_0\, V_{\rm tot}$ with $1/\lambda_{\rm tot}$, and this
ambiguity cannot be resolved within the accuracy of the BDMPS-Z formalism. We note that taking
this alternative choice, one would find for instance $\langle N_g^{\rm coh}\rangle = \frac{\lambda_{\rm tot}}{\lambda_{\rm inel}}
\left(1 - e^{-L/\lambda_{\rm tot}} \right)$. In contrast to (\ref{eq3.14}), this is smaller than 
unity for arbitrary values of $\lambda_{\rm inel}$ and $\lambda_{\rm el}$, while equation 
(\ref{eq3.14}) can be larger than unity for $\lambda_{\rm el} > \lambda_{\rm inel}$. 
In the region $\lambda_{\rm el} \ll \lambda_{\rm inel}$, for which the BDMPS-Z formalism
was derived, this difference becomes negligible.

\subsection{MC algorithms for the incoherent and totally coherent BDMPS-Z limits}
\label{sec3.2}

We consider a medium composed of scattering centers of a given density $n_0$
that provide elastic and inelastic cross sections to a projectile parton. We work within
the approximations of the BDMPS-Z formalism, that means: We neglect elastic 
scatterings of the projectile partons, since they are unimportant for gluon radiation.
And we neglect subsequent inelastic scatterings of the radiated gluons, since they are
unimportant for understanding the energy loss of the projectile parton. 

\subsubsection{MC algorithm for the incoherent BDMPS-Z limit}
\label{sec3.2.1}

We first formulate a MC algorithm that implements the BDMPS-Z formalism in the absence 
of quantum interference effects (incoherent limit). The starting point of the probabilistic evolution
is a partonic projectile that propagates on a straight line $\xi \in [0; L]$ through a medium of
path-length $L$. The interaction between projectile and medium is characterized fully in terms
of the inelastic mean free path $\lambda_{\rm inel}$ of the projectile and the elastic mean
free path $\lambda_{\rm el}$ of the radiated gluons. The dynamic evolution starts at $\xi = 0$
and it proceeds according to the following steps:

\begin{enumerate}
\item \underline{Determine whether and where the projectile undergoes its next inelastic scattering}\\
Decide with probability $1-S_{\rm no}^{\rm proj}(L)$ that a scattering occurs within the
remaining in-medium path length $L$. Here, $S_{\rm no}^{\rm proj}(L)$ is the 
probability that the projectile does not undergo any inelastic interaction within length $L$,
\begin{equation}
	S_{\rm no}^{\rm proj}(L) = \exp\left({-L/\lambda_{\rm inel}}\right)\, .
			\label{eq3.15}
\end{equation}
If no further inelastic interaction is found, then stop the dynamical evolution. 
Else, determine the distance $\xi$ to the next inelastic scattering center according to the probability density
		  \begin{equation}
		  	\Sigma(\xi) = - \frac{\dr S_{\rm no}^{\rm proj}(\xi)}{\dr\xi}
			=  \frac{1}{\lambda_{\rm inel}}\, S_{\rm no}^{\rm proj}(\xi)\, .
					\label{eq3.16}
		  \end{equation}
\item \underline{After inelastic scattering, continue propagating the projectile}\\
After an inelastic interaction at position $\xi$, the outgoing projectile has a remaining  in-medium path length $L-\xi$. 
To establish whether the projectile undergoes further inelastic interactions, 
repeat step 1 with inelastic no-scattering probability $S_{\rm no}^{\rm proj}(L-\xi)$. Reiterate this step till
no further inelastic interaction is found.
\item \underline{After inelastic scattering, propagate the produced gluon}\\
The gluon, produced in an inelastic process at position $\xi$, has a remaining in-medium path-length $L-\xi$.
Determine the number and positions of additional elastic interactions of the gluon with the medium as follows:\\
Determine whether and where the gluon undergoes its next elastic scattering,
	based on the elastic no-scattering probability 
	$S_{\rm no}^{\rm el}(L-\xi) = \exp\left[{-(L-\xi)/\lambda_{\rm el}}\right]$.
	That means, decide with probability $1-S_{\rm no}^{\rm el}(L-\xi)$ that there is another elastic scattering, and
	determine its distance $\xi'-\xi$ according to the probability density
		  \begin{equation}
		  	\Sigma(\xi'-\xi) = - \frac{\dr S_{\rm no}^{\rm el}(\xi'-\xi)}{\dr\xi'}
			=  \frac{1}{\lambda_{\rm el}}\, S_{\rm no}^{\rm el}(\xi'-\xi)\, .
					\label{eq3.17}
		  \end{equation}
Reiterate this process for each gluon till no further elastic scattering center is found.
\end{enumerate}

\noindent
According to this MC algorithm, the probability $P_{\rm inel}(m)$ for generating dynamical scattering histories with 
exactly $m$ inelastic interactions is determined by reiterating step 2 in the above algorithm,
\begin{eqnarray}
	 P_{\rm inel}(m) &=&
	 \int_{0}^L \frac{\dr x_1}{\lambda_{\rm inel}} e^{-x_1/\lambda_{\rm  inel}}
	 	 \int_{x_1}^L \frac{\dr x_2}{\lambda_{\rm inel}} e^{-(x_2-x_1)/\lambda_{\rm inel}}
	  \dots
	   \int_{x_{m-1}}^L \frac{\dr x_m}{\lambda_{\rm inel}}\, 
	   	\exp\left[- (L-x_m)/\lambda_{\rm inel} \right]
		\nonumber \\
		&=& \exp\left[- \frac{L}{\lambda_{\rm inel}} \right]\, 
		\frac{1}{m!}
		\left( \frac{L}{\lambda_{\rm inel}}\right)^m\, .
		 \label{eq3.18}
\end{eqnarray}
Since the algorithm produces exactly one gluon per inelastic interaction, $P_{\rm inel}(m)$ is the
probability for finding scattering histories with exactly $m$ produced gluons. The  average number 
of gluons per scattering history is 
\begin{equation}
	\langle N_g^{\rm MC, incoh} \rangle = \sum_{m=1}^\infty
	P_{\rm inel}(m) \, m 
	= \frac{L}{\lambda_{\rm inel}}\, ,
			\label{eq3.19}
\end{equation}
which is consistent with the corresponding incoherent limit in the BDMPS-Z formalism, see (\ref{eq3.10}). 

\subsubsection{MC algorithm in the presence of coherence effects}
\label{sec3.2.2}

Coherence effects in gluon production processes can be accounted for by modeling the production
as taking place over a finite formation
time $\tau_f$ in (\ref{eq2.20}). The incoherent limit of gluon production is then realized for the
case $\tau_f \ll \lambda_{\rm inel}\, , \lambda_{\rm el}$ and the totally coherent limit is realized for 
$\tau_f \gg L$. To decide which of these limits applies to a specific gluon production process, the
MC algorithm needs to know $\tau_f$. The dynamical determination of $\tau_f$ requires
${\bf k}$-differential information and will be discussed in the context of the ${\bf k}$-differential
algorithm in section~\ref{sec4}. As a preparatory step, we explore here the formal limits $\tau_f \to 0$
(incoherent) and $\tau_f \to \infty$ (totally incoherent) gluon radiation, and we study in these limits
 ${\bf k}$-integrated yields.  We want to devise an algorithm 
that extends naturally to a ${\bf k}$-differential version. To this end,  we should use 
information about whether we work in the totally coherent or incoherent limit  only in 
algorithmic steps in which information about $\tau_f$ would be dynamically available in the
${\bf k}$-differential version. Therefore, as long as the inelastic scattering and its kinematics
is not yet determined, the MC algorithm must still allow for the cases that the inelastic production process 
could be either incoherent or could include coherence effects. This consideration prompts us to seek an MC 
implementation that starts from selecting an inelastic process as 
in the incoherent case, based on equations (\ref{eq3.15}) and (\ref{eq3.16}). 
Coherence effects will then be included  by modifying the subsequent evolution and by reweighting the 
inelastic process that was selected with the probability of an incoherent production. Such reweighting is a 
standard Monte Carlo technique in algorithms that overestimate probabilities. We discuss now
both these elements in more detail:

\noindent
\underline{\bf Modifying the subsequent evolution:}\\
Assume that the MC algorithm has selected an inelastic process 'at $\xi$' according to (\ref{eq3.15}), (\ref{eq3.16}),
and that the formation time $\tau_f$ of the produced gluon is then found to be finite.  How should this be taken into 
account in the further probabilisitic evolution? 
The general idea is that if $\tau_f$ cannot be neglected ( $\tau_f > \lambda_{\rm inel}\, , \lambda_{\rm el}$),
then the position $\xi$ selected in (\ref{eq3.16}) cannot be interpreted as the 'point' of the gluon emission. 
Rather, we view the simulated pair of values $\xi$, $\tau_f$ as specifying 
{\it a region of extent $\tau_f$ around $\xi$}, over which  the gluon production process takes place. 
Technically, this translates into the requirement that if gluon production could have started as early as 
$\xi_{\rm init} = {\rm max}\left[\xi - \tau_f; 0\right] $, then the produced gluon is allowed to scatter elastically
from time $\xi_{\rm init}$ onwards, and not only after time $\xi$. This is a modification of step 3 of the incoherent algorithm.
Physically, it means that within this entire region between $\xi_{\rm init}$ and $\xi_{\rm init} + \tau_f$ , elastic interactions
act coherently with the inelastic one.

In the present subsection, we restrict our discussion to the totally coherent case, $\tau_f \gg L$. 
In this particular limit, irrespective of the position $\xi$ at which the MC algorithm allocates the center 
of an inelastic process, this process is delocalized over the entire in-medium path length $L$. As a consequence, 
irrespective of the choice of $\xi$, the radiated gluon can 
accumulate additional elastic interactions between $\xi_{\rm init} = 0$ and $L$.

\noindent
\underline{\bf Reweighting inelastic processes:}\\
In the incoherent case, the probability that the projectile parton undergoes one or more inelastic interactions
is given by $1 - S_{\rm no}^{\rm proj}(L)$, see  (\ref{eq3.15}). 
Each scattering center serves as an independent source of gluon production. In contrast, in the
presence of coherence effects, it is the ensemble of several scattering centers that acts effectively
as one source of gluon production. Therefore,
the factor $1 - S_{\rm no}^{\rm proj}(L)$ overestimates the probability of inelastic interactions, 
and a reweighting is needed.

To determine this reweighting factor, we observe that in the totally coherent limit of 
the BDMPS-Z formalism,  $\langle N_g^{\rm coh} \rangle (N_s)$ in  (\ref{eq3.13}) denotes the average number of gluons produced with exactly $(N_s-1)$ elastic and one inelastic interaction. The corresponding 
expression in the incoherent limit is given in (\ref{eq3.10}) and it is one power of $N_s$ larger, 
$\langle N_g^{\rm incoh} \rangle (N_s) = N_s\, \langle N_g^{\rm coh} \rangle (N_s)$. 
Therefore, the $N_s$-averaged number of emitted gluons can be obtained in the totally coherent limit, if
a gluon selected according to (\ref{eq3.15}), (\ref{eq3.16}) and having undergone
$N_s$ scatterings is accepted with probability
\begin{equation}
	w = \frac{1}{N_s}\, .
	\label{eq3.20}
\end{equation}

Based on these considerations, we propose the following MC algorithm for the totally coherent limit:

\begin{enumerate}
\item \underline{Determine whether the projectile undergoes an inelastic scattering.}\\
As in the incoherent case, use (\ref{eq3.15}) to decide with probability $1-S_{\rm no}^{\rm proj}(L)$ that a
scattering 
occurs within the in-medium path length $L$. If no inelastic interaction is found, then stop the dynamical evolution. \\
\item \underline{After inelastic scattering, continue propagating the projectile.}\\
Establish whether the projectile undergoes further inelastic interactions by searching with 
probability $1-S_{\rm no}^{\rm proj}(L-\xi)$ for further inelastic scatterings between $\xi$ and $L$. 
 \\
\item \underline{After inelastic scattering, propagate the produced gluon up to length $L$ }\\
\underline{and reweight its production probability.}\\
In the totally coherent case, the production is delocalized over the entire medium of length $L$
and therefore, all gluons undergo elastic scattering over an in-medium path length $L$.
With probability $1- w = 1- \frac{1}{N_s}$, the produced gluons are rejected.
\end{enumerate}

\subsection{Validating the proposed MC algorithms}
\label{sec3.3}
We have written MC programs that implement the algorithms proposed in sections~\ref{sec3.2.1} and ~\ref{sec3.2.2}
for the case of incoherent and totally coherent gluon production, respectively. To check that these algorithms 
reproduce the analytically known results of the BDMPS-Z formalism, we establish that
they account for the average number of gluons  produced per scattering history in both limits,
$\langle N_g^{\rm coh}\rangle$ and $\langle N_g^{\rm incoh}\rangle$. In addition, the MC algorithms allow us
to plot the average number of gluons $\langle N_g\rangle^{(\rm incoh)}_j$ and $\langle N_g\rangle^{(\rm coh)}_j$,
produced with exactly $j$ momentum transfers from the medium. Here, we test against this more differential
information. 

In the totally coherent limit, we see from equation (\ref{eq3.12}) that the expansion of $\langle N_g^{\rm coh}\rangle(N_s)$
to order $N_s$ involves gluon radiation terms with exactly $N_s$ momentum transfers. As a consequence,
the average number of gluons produced with exactly $j$ momentum transfers is given by
\begin{equation}
  \langle N_g\rangle^{(\rm coh)}_j = \langle N_g^{\rm coh}\rangle(N_s=j)\, .
  \label{eq3.21}
\end{equation}
The analogous identification of orders in the opacity expansion with number of momentum transfers does not hold
in the incoherent limit. As one sees for instance from equation (\ref{eq3.9}), the second order receives
contributions from gluons that were produced either with one single momentum transfer (these are the terms
$R({\bf k},{\bf q}_1)$)  or with two momentum transfers (these are the terms $R({\bf k}+{\bf q}_1,{\bf q}_2)$).
To identify all contributions with a fixed number of momentum transfers, we write the incoherent limit of the 
BDMPS-Z formalism as a series
\begin{eqnarray}
	\omega \frac{\dr I}{\dr\omega\, \dr{\bf k}} \Bigg\vert^\text{incoh}
	&=&  {\alpha_s\over \pi^2}\, C_R\, \exp\left(-n_0\, L\, V_{\rm tot} \right)\, 
		\sum_{N=1}^\infty \frac{1}{N!}\, \left(n_0\, L\, V_{\rm tot} \right)^{N-1}\, 
		n_0\, L\, \int_{{q}_1}  R({\bf k},{\bf q}_1)
		\nonumber \\
		&& + {\alpha_s\over \pi^2}\, C_R\, \exp\left(-n_0\, L\, V_{\rm tot} \right)\, 
		\sum_{N=2}^\infty \frac{1}{N!}\, \left(n_0\, L\, V_{\rm tot} \right)^{N-2}\, 
		(n_0\, L)^2\, \int_{{q}_1}  \int_{{q}_2} R({\bf k}+{\bf q}_2,{\bf q}_1)
		\nonumber \\
		&& ...
		 \label{eq3.22}
\end{eqnarray}
Here, contributions involving the radiation term $R\left({\bf k}+\sum_{i=2}^j {\bf q}_i,{\bf q}_1\right)$
denote gluon production processes with $j$-fold scattering (i.e. with $(j-1)$-fold elastic scattering). 
Integrating formally over phase space, one finds that the average number of such gluons per event, 
produced with $j$-fold scattering, can be expressed in
terms of complete and incomplete $\Gamma$-functions,
\begin{eqnarray}
	\langle N_g\rangle^{(\rm incoh)}_j &=& 
		 \exp\left(- \frac{L}{\lambda_{\rm el}} \right)\,  \sum_{N=j}^\infty
		\frac{1}{N!} 
		\left( \frac{L}{\lambda_{\rm el}} \right)^N
		 \left( \frac{\lambda_{\rm el}}{\lambda_{\rm inel}} \right)
		 \nonumber \\
		 &=& \frac{L}{\lambda_{\rm inel}} 
		 \frac{\Gamma(j) - \Gamma(j,\frac{L}{\lambda_{\rm el}})}{\frac{L}{\lambda_{\rm el}}\,\Gamma(j)}
		 		\label{eq3.23}
\end{eqnarray}
One can check that the average number of incoherently produced gluons is again given 
by 
\begin{eqnarray}
	\langle N_g\rangle^{(\rm incoh)} &=& \sum_{j=1}^\infty \langle N_g\rangle^{(\rm incoh)}_j 
	 = \frac{L}{\lambda_{\rm inel}}\, . 
	 \label{eq3.24}
\end{eqnarray}
%
\begin{figure}[h]
\begin{center}
\includegraphics[height=12.0cm, angle=0]{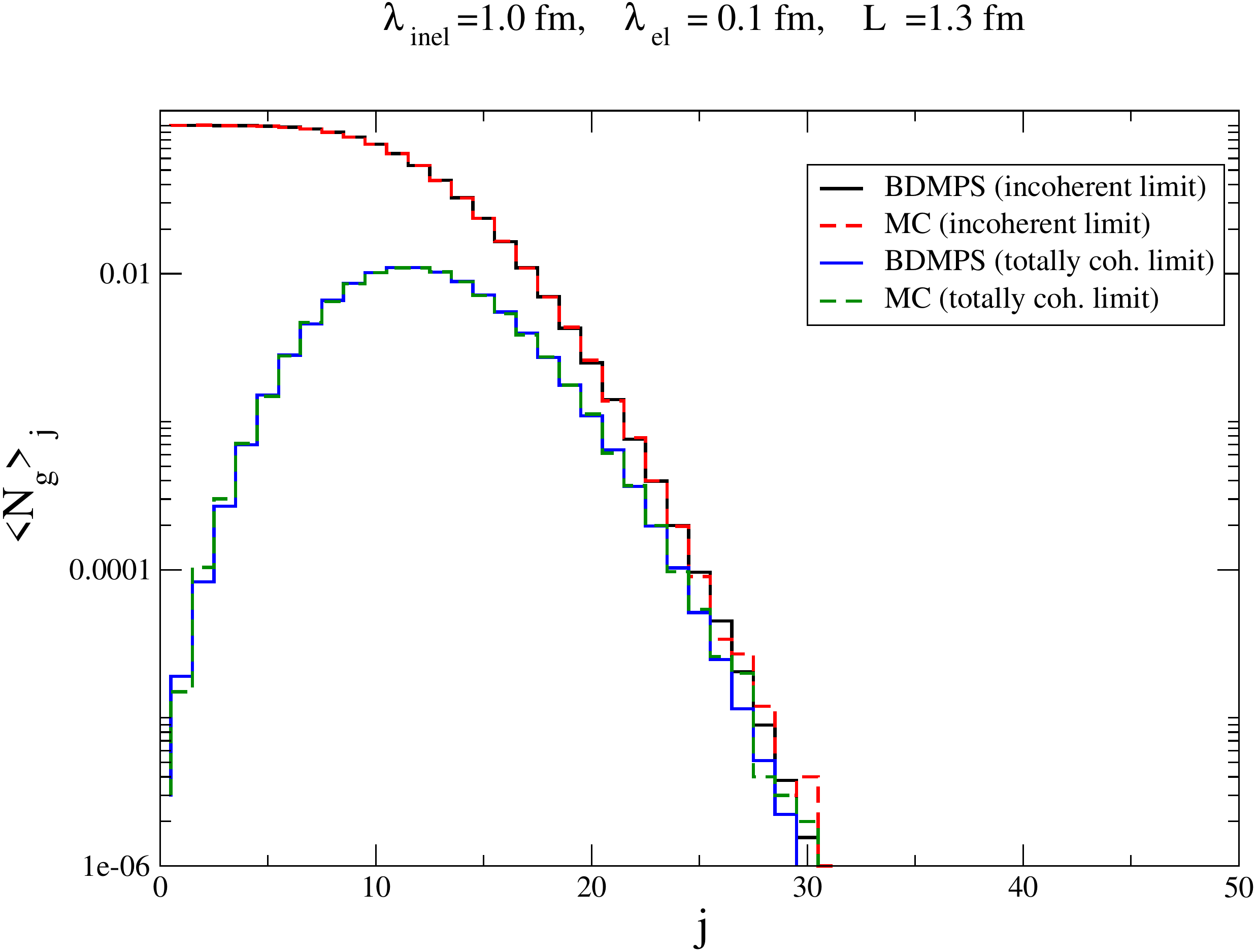}
\caption{The average number of gluons, produced with exactly $j$ momentum transfers
from the medium. Results are shown for some arbitrary choice of inelastic and elastic 
mean free path, and total in-medium path length $L$. Analytical results of the BDMPS-Z
formalism are compared to MC simulations in the totally 
coherent and incoherent limits.}
\label{fig1}
\end{center}
\end{figure}
%
Fig.~\ref{fig1}  shows analytical results for $\langle N_g\rangle^{(\rm incoh)}_j$ and $\langle N_g\rangle^{(\rm coh)}_j$,
compared to output of the MC programs implementing the algorithms of section~\ref{sec3.2.1} and ~\ref{sec3.2.2}. 
We have tested that the proposed algorithms reproduce the results of the BDMPS-Z formalism
for a broad range of values of the inelastic and elastic mean free path and for different in-medium path lengths $L$.  
The differences between analytical results and MC simulations could always be decreased arbitrarily by increasing 
sufficiently the number of events in the MC simulation. This establishes that the proposed algorithms implement the 
BDMPS-Z formalism for ${\bf k}-$ and $\omega$-integrated quantities.

\section{A ${\bf k}$- and $\omega$-differential MC algorithm in the totally coherent and incoherent BDMPS-limits}
\label{sec4}
In the previous section, we have shown how the coherent and incoherent limits of the phase space integrated 
average number of radiated gluons $\int \dr\omega \int \dr{\bf k}\, \frac{\dr I}{\dr \omega\, \dr{\bf k}}$ can be
simulated in
a probabilistic MC algorithm. In the present section, we extend these algorithms to 
a simulation of the differential gluon distribution $\frac{\dr I}{\dr\omega\, \dr{\bf k}}$.

The basic building block for the differential distribution $\int \dr\omega \int \dr{\bf k}\, \frac{\dr I}{\dr \omega\,
\dr{\bf k}}$ 
is the inelastic interaction of the projectile with a single scattering center. According to eqs. (\ref{eq2.11})
and (\ref{eq2.16}), the corresponding inelastic cross section is
\begin{equation} 
	 \frac{\dr\sigma_{\rm inel}}{\dr\omega\, \dr{\bf q}\, \dr{\bf k}}  
	 = {\alpha_s\over \pi^2}\, 
   C_R\, 
    \frac{1}{\omega} \frac{1}{(2\pi)^2}\,
   \vert A({\bf q})\vert^2 
    \frac{{\bf q}^2}{{\bf k}^2\left({\bf k}+{\bf q} \right)^2 }\, .
		\label{eq4.1}
		\end{equation}
We seek a MC algorithm that interpolates between the coherent and incoherent limits by
treating all momentum transfers during the formation time of a gluon as coherent, and all 
scatterings outside the formation time as incoherent. Such an algorithm must keep track of 
the kinematics of the scatterings, and it must account dynamically for changes in the formation 
time. We propose an algorithm that as criterion for decoherence of the gluon
requires the relative phase of the radiated gluon
\begin{equation}
 \varphi(\Delta z) = \left \langle \frac{k_\perp^2}{2 \omega}\, \Delta z \right \rangle
 \label{eq4.2}
\, ,
\end{equation}
to become unity. More precisely, we observe that the interference factor (\ref{eq2.22}) 
extracted from the BDMPS-Z formalism is best approximated by a $\Theta$-function of the 
form~\footnote{The interference factor $f(x=L\, Q_1) = 2\left(1-\cos x\right)/x^2$ decreases
continuously from $f(0) = 1$ to $f(2\pi) = 0$, and it oscillates for larger values of $x$
with rapidly decreasing amplitude $\propto 1/x^2$. One finds $\int_0^{2\pi} f(x)\, \dr x = 2.84$
and  $\int_0^{\infty} f(x)\, \dr x = \pi$. Choosing the step of the $\Theta$-function at $x=3$
appears to be a fair representation of the main quantitative features of $f(x)$, but we do not
have an a priori argument for excluding slightly different values. }
\begin{equation}
	n_0^2 \frac{1-\cos (L\, Q_1)}{Q_1^2} \approx  \frac{\left(n_0\, L\right)^2}{2}
		\Theta\left( 3 - L\, Q_1\right)\, .
		\label{eq4.3}
\end{equation}
Therefore, we define the formation time by the condition 
\begin{equation}
 \varphi(\tau_f) \equiv 3\, .
 \label{eq4.4}
 \end{equation}
 We first discuss in section~\ref{sec4.1} the inputs and approximations of (\ref{eq4.1}) that simplify
an MC implementation. We then specify a MC algorithm before discussing how some of these
approximations can be relaxed.  

\subsection{Inputs and approximations in the proposed MC algorithm}
\label{sec4.1}

In the study of parton energy loss models and the BDMPS-Z formalism, a standard parametrization 
of elastic scattering cross sections is in terms of a Yukawa potential with a screening mass $\mu$,
\begin{equation}
	\vert A({\bf q})\vert^2 \propto \frac{1}{\left( {\bf q}^2 + \mu^2\right)^2}\, .
	\label{eq4.5}
\end{equation}
In the following, we work with this ansatz for 
$\mu \in \left[\unit[100]{MeV}; \unit[1]{GeV}\right]$.  

In equation (\ref{eq4.1}), the inelastic cross section for a single incoherent scattering 
factorizes into the product of the elastic cross section
and a radiation term. The term $R({\bf k}; {\bf q})$ specifies how gluons produced with 
energy $\omega$ are distributed in transverse phase space prior to undergoing subsequent interactions. 
What matters for the decoherence of the gluon is its relative momentum with respect to the outgoing parent 
parton. If the final transverse momentum of the gluon is build up by many interactions with the medium,
then the precise distribution of the transverse momentum at the inelastic interaction can be expected to
be unimportant. Moreover, even if there are not many interactions with the medium, the transverse momentum 
of the gluon at the inelastic interaction will be set by the recoil received by the medium. These
considerations prompt us to adopt the following approximation that  simplifies the numerical implementation 
significantly 
\begin{eqnarray} 
 \frac{\dr \sigma_{\rm inel}}{\dr\omega\, \dr{\bf q}\, \dr{\bf k}} 
 &=&
{\alpha_s\over \pi^2}\, 
   C_R\, 
    \frac{1}{\omega} \frac{1}{(2\pi)^2}\,
   \vert A({\bf q})\vert^2 
    \frac{{\bf q}^2}{{\bf k}^2\left({\bf k}+{\bf q} \right)^2 }\nonumber\\
&& \approx  f_{\rm prop}\, 
	\frac{\dr\sigma_{\rm el}}{\dr{\bf q}} \frac{\alpha_s\, C_R}{\omega}\,
      (2\pi)^2\, \delta^{(2)}\left( {\bf k}-{\bf q} \right)\, .
	            \label{eq4.6}
\end{eqnarray}
In section~\ref{sec5}, we shall provide numerical evidence that the approximation (\ref{eq4.6})
is sufficient for a quantitative MC implementation of the BDMPS-Z formalism. With the help
of (\ref{eq4.6}), the total inelastic cross section simplifies to
\begin{equation}
	\sigma_\text{inel} =  f_{\rm prop}\,  \sigma_\text{el}  \, \alpha_s\, C_R\, \log\left(\omega_{\rm
max}/\omega_{\rm min} \right) \, .
	\label{eq4.7}
\end{equation}
Here, we have considered gluon radiation in the range $\omega \in \left[ \omega_{\rm min}; \omega_{\rm max}\right]$.
We note that the first line of (\ref{eq4.6}) needs to be regularized, since
the integral over $R({\bf k}; {\bf q})$ is infrared divergent. Performing the integral over 
$R({\bf k}; {\bf q})$ with an infrared cut-off $\epsilon$ around  ${\bf k} = 0$ and ${\bf k} = {\bf q}$, 
one finds $f_{\rm prop} = \frac{2}{\pi} \left[\log \left( \mu^2/\epsilon \right) + {\rm const.} \right]$. 
In our MC algorithm, the infra-red regulator $\epsilon$ will not appear. Rather,  for one arbitrary 
choice of model parameters, we shall adjust $f_{\rm prop}$ such that the BDMPS result for the
average parton energy loss is reproduced with the correct norm. For all other choices of model
parameters, $f_{\rm prop}$ is then kept fixed and the MC algorithm generates normalized results. 
What can be said a priori about the numerical value of $f_{\rm prop}$ is that there is no physical
reason for choosing an infrared regulator $\epsilon$ that is much smaller than the momentum
scale $\mu$. Therefore, the logarithm $\log \left( \mu^2/\epsilon \right) $ should not be large,
and $f_{\rm prop}$ should be of order unity. We shall confirm this expectation in section~\ref{sec5}. 

We pause to comment on this approximation from a wider perspective: The BDMPS-Z formalism
(\ref{eq2.2}) does not depend on total elastic and inelastic cross sections, but only on the dipole
cross section (\ref{eq2.4}) that does not require regularization since it is differential in configuration 
space. However, the opacity expansion of (\ref{eq2.2}) rearranges this formalism in a series
that does contain total phase-space integrated quantities. To arrive at a probabilistic implementation, 
we have assigned to some terms in the opacity expansion of (\ref{eq2.2})
the natural physical meaning of elastic and inelastic cross sections and of mean free paths 
(see eqs.~(\ref{eq3.6}) and (\ref{eq3.8})). This can only be done with the help of approximations
and regularizations that are not explicit in the BDMPS-Z formalism (\ref{eq2.2}). For instance,
the identification of phase-space integrated expressions of the opacity expansion with rational
functions of mean free paths (such as e.g. eq.~(\ref{eq3.9})) is strictly speaking a physically motivated 
assignment rather than an analytically derived fact, since the transverse momentum integrals are 
infra-red divergent.  
The crucial test for the MC implementation is then that physical results do not depend on the 
regularization prescriptions employed and that they account quantitatively for the BDMPS-Z formalism (\ref{eq2.2}).
That this is so will be demonstrated in section~\ref{sec5}.

\subsection{A ${\bf k}$- and $\omega$-differential MC algorithm interpolating between the incoherent and totally
coherent BDMPS-limits}
\label{sec4.2}

\begin{enumerate}

 \item \underline{Initialisation}\\ Set remaining path length of the
projectile to total path length, $L_\text{proj}=L$.

\item \underline{Determine whether and where the projectile undergoes its next inelastic scattering}\\
This step is implemented as described by equations (\ref{eq3.15}), (\ref{eq3.16}) and 
accompanying text. If an inelastic scattering is generated at position $\xi$, then the remaining
path length of the projectile is set to $L-\xi$. The produced gluon is propagated further according to
the step 3 below. The algorithm repeats step 2 till no further inelastic scatterings are found in the 
remaining path length. 

 \item \underline{Kinematics of gluon emission and dynamical evolution of formation time}\\ 
 In the BDMPS-Z formalism, the gluon energy is distributed according to $1/\omega$. From this 
 distribution, the gluon energy is generated. The initial transverse momentum of the gluon is 
generated from the distribution $\vert A({\bf k})\vert^2$; the initial gluon phase is taken to vanish,
$\varphi=0$;  the number of momentum
transfers to the gluon is set to $N_\text{s}=1$, and the initial formation time is determined according to 
\begin{equation}
 \tau_\text{f} = (1-\varphi) \frac{2 \omega}{{\bf k}^2} \, .
 \label{eq4.8}
\end{equation}
Then set the remaining gluon path length to the total path length, $L_\text{gluon}=L $, and check for
further elastic momentum transfers within the formation time:
\begin{itemize}
\item With probability $1-S_{\rm no}^{\rm el}(\min(\tau_\text{f}, L_\text{gluon}))$ there is one more scattering.
Determine the distance $\Delta L$ to the scattering centre and update the
path length, $L_\text{gluon}=L_\text{gluon}-\Delta L$, and the gluon phase
\begin{equation}
 \varphi = \varphi + \frac{{\bf k}^2}{2 \omega} \Delta L \, .
 \label{eq4.9}
\end{equation}
Determine the momentum transfer $\mathbf{q}_{N_\text{s}}$ from the scattering
centre according to $|A(\mathbf{q}_{N_\text{s}})|^2$, set the transverse momentum 
of the gluon  to ${\bf k} = \sum_{i=1}^{N_\text{s}} \mathbf{q}_i$, and set 
$N_\text{s} = N_\text{s} + 1$. 
Iterate this point until no further scattering is found. 
 \item With probability $S_{\rm no}^{\rm
el}(\min(\tau_\text{f}, L_\text{gluon}))$ there is no further scattering. 
Continue with point~4.
 \end{itemize}
\item \underline{Reweight the gluon production probability, and propagate gluons further.}\\
The gluons simulated in point~3 are trial gluons that have been selected with an overestimated
production probability. Reweighting is needed to correct for this overestimate. If a trial gluon is 
generated with $N_\text{s}$ scattering centers within its formation time, then
\begin{itemize}
\item With probability $1 - 1/N_\text{s}$, reject the gluon from the sample.
\item With probability $1/N_\text{s}$, accept the gluon as part of the scattering history. 
Determine the end of the formation process of the gluon by localizing a formation time interval
$\tau_f$ in an arbitrary fashion around the initial production point $\xi$. Then determine further
elastic momentum transfers to the gluon within the in-medium path length after formation has
been completed. (This last step is needed only for the simulation of ${\bf k}$-differential 
spectra.)
\end{itemize}
\item \underline{Accept only medium-induced gluons}\\
To reproduce the radiation spectrum (\ref{eq2.2}) for $\xi_0=0$,
accept only gluons that are fully formed prior to leaving the medium. 
\end{enumerate}

It is a consequence of the approximation (\ref{eq4.6}), that the gluon  transverse momentum
is build up identically in the coherent and incoherent case. We note as an aside, that it is 
possible to amend the above proposal such that it does not invoke the approximation (\ref{eq4.6}).
To do so, one has to start from the observation that a gluon produced with $N_\text{s}$
coherently acting scattering centers is produced according to the probability 
 \begin{equation}
 \int_{{\bf k}, \omega} \left( \prod_{i=1}^{N_\text{s}} \left( \int  \frac{\dr{\bf q}_i}{(2\pi)^2} 
	\frac{\vert A({\bf q}_i)\vert^2}{V_{\rm tot}} \right) \right)\, 
	{\alpha_s\over \pi^2}\, \frac{C_R}{\omega}
	R\left( {\bf k},  \sum_{j=1}^{N_\text{s}} {\bf q}_j \right) \, .
	\label{eq4.10}
 \end{equation}
In our simplified algorithm, this expression is approximated by a factor 
$\lambda_{\rm el}/\lambda_{\rm inel}$, and $\lambda_{\rm inel}$ specifies the probability 
with which an inelastic scattering occurs. There are standard reweighting techniques that
would allow one to overestimate the probability of inelastic interaction and to then correct
it to the factor (\ref{eq4.10}). In the present work, we did not exploit this numerically more
demanding procedure, and we did not find any indication that such a procedure is needed
to reproduce quantitatively the BDMPS-Z formalism (\ref{eq2.2}). 

The idea that the concept of formation time plays a central role in the probabilistic implementation of
medium-induced gluon radiation has been formulated previously. However, in our effort to arrive at a 
quantitatively reliable, probabilistic,  formation time based formulation of the BDMPS-Z formalism, 
we had to overcome several conceptions that were naively assumed at least by us, but possibly 
also by others. In particular, a MC formulation that selects gluon production processes according to
an incoherent inelastic scattering probability overestimates gluon production in the presence
of interference effects. A quantitatively reliable implementation must correct for this overestimate, and
the algorithm proposed here is, as far as we know, the first one that does so. On general grounds, one
expects that this feature is not specific for the BDMPS-Z formalism, but persists in more complete 
formulations of radiative parton energy loss. 
Secondly, it turns out
that the BDMPS-Z formalism cannot be implemented exactly in a formulation that interprets formation
times as deadtimes for subsequent gluon production. Technically, this can be seen from the form of the
average number of radiated quanta $\langle N_g\rangle_j$ as a function of the number of active scattering 
centers $j$, discussed in subsection~\ref{sec3.3}. (Formulations based on a dead time interpretation would
lead to expressions for $\langle N_g\rangle_j$ that contain terms $\propto \lambda_{\rm inel}$ in the
arguments of exponentials.) That formation times are not dead times for subsequent gluon production
could have been expected
on the simple ground that the BDMPS-Z formalism is based on a multiple scattering calculation
with only one gluon in the final state and therefore cannot account for the destructive interference between
different gluons. It remains to be seen whether this feature persists in more complete analytical calculations 
of medium-induced gluon emission.

\section{Numerical results on the gluon energy distribution}
\label{sec5}
The MC algorithm of section~\ref{sec3} and \ref{sec4} is tailored to provide a probabilistic 
implementation of the opacity expansion of (\ref{eq2.2}). At fixed order in opacity, terms in (\ref{eq2.2})
can be pictured as arising from interactions of the partonic projectile with a fixed number of scattering
centers. This {\it discrete} picture of the medium lends itself naturally to a MC implementation, and the proposed
algorithm reproduces the analytically known distribution in the number of scattering centers, see Fig.~\ref{fig1}.

In contrast, in the multiple soft scattering limit of (\ref{eq2.2}), information about the discrete structure of the medium
is lost. This limit is obtained from a saddle point approximation of the path integral in (\ref{eq2.2}), setting $n\, 
\sigma({\bf r}) = \frac{1}{2} \hat{q}\, {\bf r}^2$ . In this approximation, the BDMPS-Z transport coefficient $\hat{q}$
characterizes the average transverse momentum squared, transferred from the medium to the projectile per unit 
path length. The medium can be pictured as providing for the projectile a {\it continuous} transverse color field whose
strength is characterized by $\hat{q}$. 

Here, we shall compare results of the proposed MC algorithm to the BDMPS-Z multiple 
soft scattering approximation of (\ref{eq2.2}) according to which the energy distribution (\ref{eq2.2}) of 
gluons emitted from a highly energetic projectile shows the characteristic $1/\sqrt{\omega}$-dependence of the 
non-Abelian Landau-Pomeranchuk-Migdal effect, 
\begin{eqnarray}
    \omega \frac{\dr I}{\dr\omega} \simeq 
           \frac{2\alpha_s C_R}{\pi} 
           \left\{
           \begin{array}[]{l} 
           \sqrt{\omega_c / 2\, \omega}  \qquad  \hbox{for}\quad  \omega \ll \omega_c
              \\ \frac{1}{12} 
                  \left(\frac{\omega_c}{\omega}\right)^2 
                  \qquad  \hbox{for}\quad  \omega \gg \omega_c
                  \end{array} \right. \, .
   \label{eq5.1}
\end{eqnarray}
This  $1/\sqrt{\omega}$-spectrum is cut-off due to formation time effects at a characteristic gluon energy 
$\omega_c = \frac{1}{2} \hat{q}\, L^2$. Integrating $\omega \frac{\dr I}{\dr\omega}$, one finds 
the average parton energy loss 
\begin{equation}
	\Delta E = \frac{1}{4} \alpha_s\, C_R\, \hat{q}\, L^2\, , \qquad \hbox{for}\quad 
	    L< L_c\equiv \sqrt{\frac{{2}\omega_{\rm max}}{\hat q}}\, .
	\label{eq5.2}
\end{equation}
Here, the critical path length $L_c$ is the maximal coherence length, which occurs for the
maximal kinematically allowed gluon 
energy $\omega_{\rm max}$ (typically taken to be the projectile energy $E_{\rm proj}$). 
For lengths $L > L_c$, one expects hence 
that different regions of the medium act incoherently to gluon production and that
$\Delta E(L)$ increases linearly with $L$. The differential distribution (\ref{eq5.1}) continues to show
the characteristic coherence effects for $L>L_c$, since each gluon entering this distribution was 
produced coherently over a distance $\tau_f$ that depends on $\omega$. 

\subsection{Multiple soft scattering limit in the MC algorithm} 
\label{sec5.1}
To realize the multiple soft scattering approximation in the
MC algorithm, we ensure first that there are many elastic interactions per inelastic mean free path. Hence, we shall 
work in the limit 
\begin{equation}
	\lambda_{\rm el} \ll \lambda_{\rm inel}\, .
	\label{eq5.3}
\end{equation}
Moreover, we ensure that all elastic interactions are soft by cutting off the power-law tails of the Yukawa scattering
potential (\ref{eq4.4}) at  $\vert {\bf q}\vert  = 2\, \mu$,
\begin{equation}
	\vert A({\bf q})\vert^2 
		\quad \longrightarrow \quad \vert A({\bf q})\vert^2 \Theta\left(2\mu - \vert {\bf q}\vert \right)\, .
	\label{eq5.4}
\end{equation}
This approximation in the MC algorithm can be shown to correspond on the analytical side to a saddle 
point approximation of the path integral (\ref{eq2.2}) by writing in (\ref{eq2.4}) 
$\sigma({\bf r}) \propto \hat{q}\, {\bf r}^2$.

The soft multiple scattering approximation of (\ref{eq2.2}) and the average parton energy loss (\ref{eq5.2})
are functions of $\alpha_s\, C_R$ and for $\hat{q}$, which are not input parameters of the MC simulation.
Instead, one specifies for MC simulations the elastic and inelastic mean free paths, and 
the average transverse momentum transfer $\mu$ in the elastic scattering
cross section. To express the BDMPS energy loss formula in terms of these input parameters, we 
rewrite  the strong coupling constant with the help of eq.~(\ref{eq4.7}),
\begin{equation}
	\alpha_s\, C_R = \frac{\lambda_{\rm el}}{\lambda_{\rm inel}} 
	\frac{1}{f_{\rm prop}\, \log\left(\omega_{\rm max}/\omega_{\rm min} \right)}\, .
	\label{eq5.5}
\end{equation}
From the MC simulation, we determine the event averaged squared transverse momentum  
$\langle {\bf q}^2\rangle$ transfered from the medium to a radiated 
gluon per unit path length $L_p$,
\begin{equation}
	q_{\rm eff} \equiv \frac{\langle {\bf q}^2\rangle}{L_p}\, .
	\label{eq5.7}
\end{equation}
We then define operationally~\footnote{\label{foot1}
In a simplified scenario in which a fixed ${\bf k}^2 = \mu^2$ is transferred per mean free path $\lambda_{\rm el}$
from the medium to a gluon, the MC algorithm will accumulate within a length $L = n\, \lambda_{\rm el}$
a gluon phase $\varphi \approx \frac{1}{2\omega} \sum_{j=0}^{n-1} j\, \mu^2\, \lambda_{\rm el}
\simeq \frac{1}{2\omega}\, \frac{q_{\rm eff}\, L^2}{2}$. This phase differs by a factor 2 from the standard 
analytical pocket estimate $\varphi = \frac{\langle {\bf k}^2 \rangle}{2\omega}\, L \simeq 
\frac{1}{2\omega}\, \hat{q}\, L^2$. The reason is that the squared transverse momentum
$\langle {\bf k}^2 \rangle_{\Delta L}$ accumulated between $L-\Delta L$ and $L$, can contribute
to $\varphi$ only with $\langle {\bf k}^2 \rangle_{\Delta L}\, \Delta L / 2\omega$
and not with $\langle {\bf k}^2 \rangle_{\Delta L}\, L / 2\omega$. 
This illustrates that pocket formulas for $\varphi$ (and a fortiori for $\omega_c$ and $L_c$) should
not be expected to provide numerically accurate prefactors but identify the parametric dependencies only.} 
\begin{equation}
  \hat{q} = q_{\rm eff}\, .
\end{equation}
In general, $\mu^2/\lambda_{\rm el}$ would be a poor approximation of $q_{\rm eff}$, but 
for the particular choice of soft scattering centers (\ref{eq5.4}) regulated at $\vert {\bf q}\vert = 2\, \mu$,
$\langle {\bf q}^2\rangle = \mu^2$ and $q_{\rm eff}$ agree with $\mu^2/\lambda_{\rm el}$. We can now
express the BDMPS parton energy loss formula (\ref{eq5.2})
in terms of input parameters of the proposed MC algorithm, 
\begin{equation}
	\Delta E = \frac{1}{4}\, \frac{1}{f_{\rm prop}\, \log\left( E_{\rm proj} / \omega_{\rm min}\right)} 
	 \frac{\lambda_{\rm el}}{\lambda_{\rm inel}} q_{\rm eff}\, L^2\, .
	\label{eq5.8}
\end{equation}
It is this form of the BDMPS parton energy loss formula that we test in the MC studies presented in this
section. 
%
\begin{figure}[h]
\begin{center}
\includegraphics[height=10.0cm, angle=0]{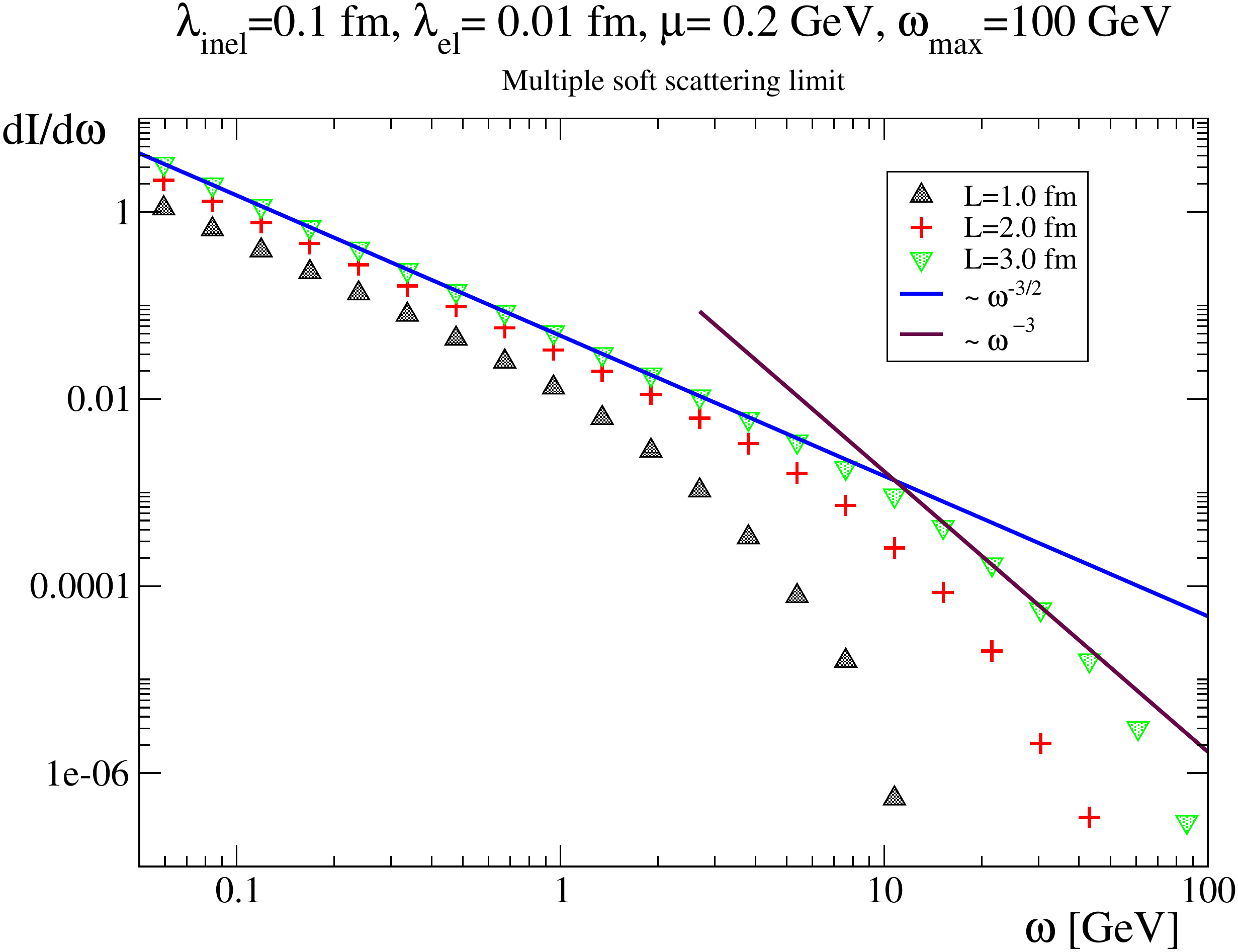}
\caption{
The spectrum of medium-induced gluons as a function of gluon energy $\omega$ for different in-medium
path lengths $L$. To compare with the soft multiple scattering limit, results have been calculated with extreme
choices of elastic and inelastic mean free paths.}
\label{fig2}
\end{center}
\end{figure}

In the following subsections~\ref{sec5.2} and ~\ref{sec5.3}, we explore the proposed MC algorithm
for values $\lambda_{\rm el} \simeq O(10^{-1})\, \lambda_{\rm inel}$ that realize the multiple scattering
approximation (\ref{eq5.3}). We note that the strong coupling constant in (\ref{eq5.3}) is proportional
to $\lambda_{\rm el}/\lambda_{\rm inel}$; moreover, it decreases with a large logarithm 
$1/\log\left(\omega_{\rm max}/\omega_{\rm min} \right) \simeq O(10^{-1})$. 
(Unless stated otherwise, the numerical results in this section are 
for $\omega_{\rm max} = 100$ GeV and $\omega_{\rm min} = 50$ MeV.) 
{\it As a consequence, the numerical values
for the average energy loss presented in the next subsections ~\ref{sec5.2} and ~\ref{sec5.3}
will be typically a factor $10^{-2} $ lower than realistic values, since they have been obtained with an 
artificially low strong coupling constant.} It is only by relaxing the multiple soft scattering approximation (\ref{eq5.3})
that realistic values of the strong coupling strength can be implemented in the present MC algorithm. This will
be done in section~\ref{sec5.4}.

%
\begin{figure}[t]
\begin{center}
\includegraphics[height=10.0cm, angle=0]{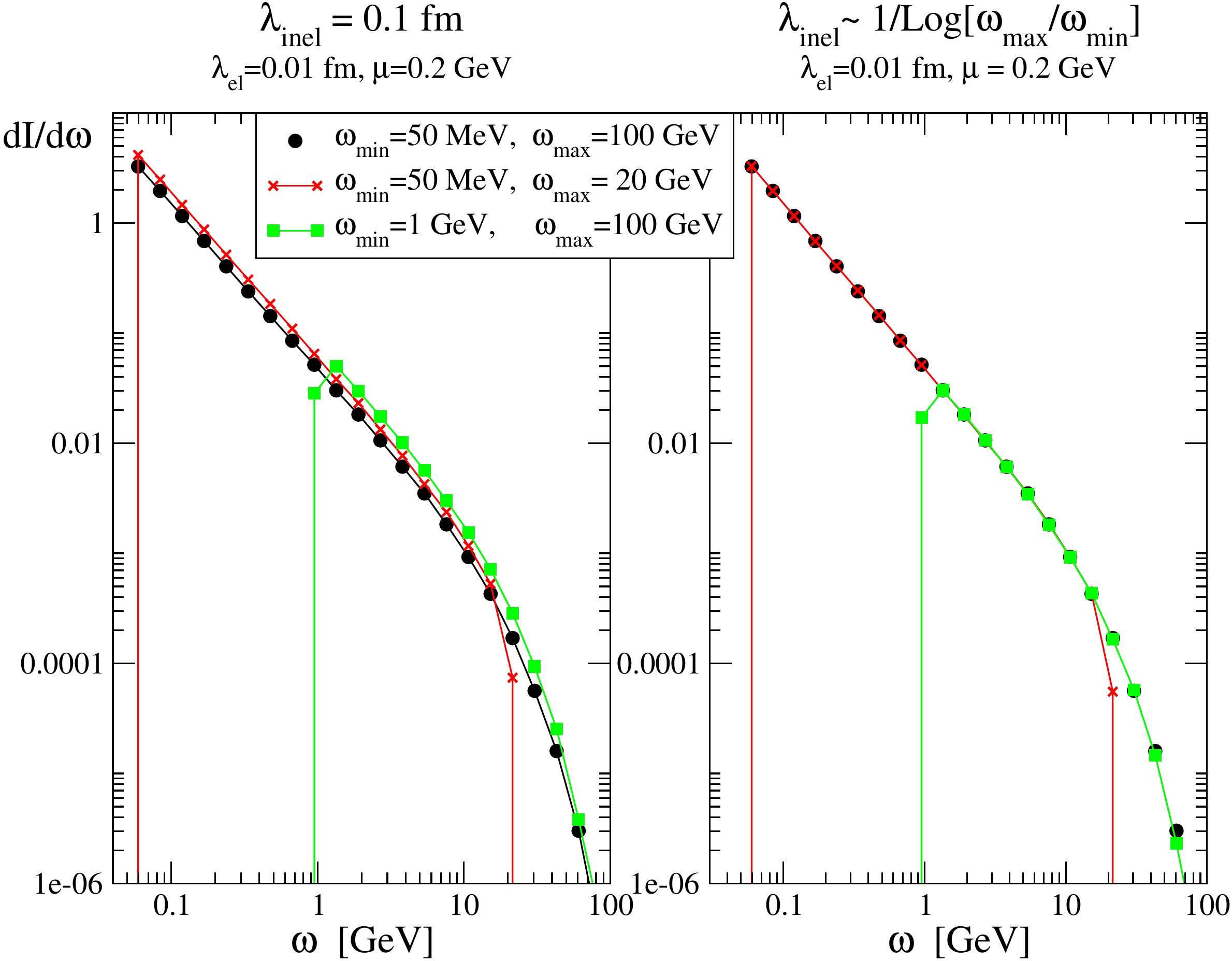}
\caption{Cut-off insensitivity of the MC algorithm. 
For rescaled inelastic mean free path $\lambda_{\rm inel}$, varying the IR and UV regulators of the 
inelastic cross section does not affect the physics results of the MC simulations, but is limited to changing the
kinematic range within which these physics results are generated.}
\label{fig3}
\end{center}
\end{figure}

\subsection{MC results of the gluon energy distribution and control of cut-off dependence}
\label{sec5.2}
Fig.~\ref{fig2} shows the medium-induced gluon spectrum for a projectile parton propagating 
through a medium of path length $L$. These and the following results were obtained for MC simulations
of $N_{\rm evt} = 10^6$ events. For sufficiently large in-medium path length $L$, the 
spectrum $\omega \frac{\dr I}{\dr\omega}$ 
approaches the characteristic $1/\sqrt{\omega}$-dependence expected for the non-abelian LPM 
effect. This dependence may be understood also by the following parametric argument: In the 
incoherent limit, gluon production on a single scattering center results in a spectrum
$\propto 1/\omega$. Coherence effects imply that the number $N_{\rm coh}$ of scattering centers located within the
formation time of the gluon act as one single effective scattering center. The resulting gluon spectrum is 
$\propto \frac{1}{N_{\rm coh}\, \omega}$. The average number of coherently acting scattering centers is proportional
to the average formation time, and this average formation time should satisfy 
$t_{\rm coh} \propto \frac{\omega}{\hat{q}\, t_{\rm coh}}$.
As a consequence, $N_{\rm coh} \propto t_{\rm coh} \propto \sqrt{\omega}$ and therefore coherence effects 
change the gluon energy spectrum by one factor $\sqrt{\omega}$.

For sufficiently small in-medium path length $L$ or sufficiently large projectile energy $\omega_{\rm max}$, 
the formation of gluons of high energy $\omega$ is suppressed since 
their formation time becomes comparable to the entire in-medium path length. Parametrically,
this suppression is expected to set in at a characteristic gluon energy $\omega_c = \frac{1}{2} \hat{q}\, L^2$,
that takes values of $\omega_c = $ 10, 40 and 90 GeV respectively for the in medium path lengths 
$L =$ 1, 2 and 3 fm explored in Fig.~\ref{fig2}. 
We note as an aside that in the limit $\omega_c\, L \to \infty$, 
the expression (\ref{eq2.2}) reduces to the BDMPS limiting result 
$\frac{\dr I}{\dr \omega} \propto \log \left| \cos\left(\frac{1+i}{\sqrt{2}}\frac{\omega_c}{\omega}\right)\right|$.
Numercial inspection of this limiting case reveals that the transition from the small-$\omega$ to the 
large-$\omega$ behavior of (\ref{eq5.1}) occurs at values that are a factor $\sim 3-5$ smaller than $\omega_c$.
This is quantitatively consistent with the location of this transition region in Fig.~\ref{fig2}, and it further
illustrates the comment in footnote~\ref{foot1}. Furthermore, a
lower value for the transition energy was also found
in~\cite{Salgado:2003rv,Salgado:2003gb}. We conclude that the proposed MC
algorithm reproduces the
$\omega^{-3/2}$-dependence of the BDMPS-Z formalism for soft gluon production up
to the expected 
scale which is of order $\omega_c$. For higher gluon energies, one observes a steeper 
$\omega$-dependence, consistent with the BDMPS-Z formalism, but
one finds some deviations from the $\omega^{-3}$-dependence of (\ref{eq5.1})  for realistic projectile energies.  
Since gluon energies $\omega \gg \omega_c$ are known to be numerically unimportant in the BDMPS-Z
formalism, these deviations will turn out to be negligible for the following.

We now turn to an issue that  is crucial for the predictive power of a MC algorithm, namely that the 
physics results of the algorithm are insensitive to the numerical choices of  IR and UV regulators,
though various intermediate steps in the algorithm may depend on the choice of such regulators. To
be specific, the MC algorithm selects  inelastic interactions with a probability 
$1 - \exp\left( -L/\lambda_{\rm inel}\right)$ that depends on the total inelastic cross section.
This cross section (\ref{eq4.7}) depends explicitly on IR- and UV regulators $\omega_{\rm min}$ 
and $\omega_{\rm max}$. 
The physics output will still be insensitive to these regulators if the dependence of the total inelastic
cross section on phase space available for radiation is respected in the MC implementation.
Technically, this is achieved in the present algorithm by rescaling $\lambda_{\rm inel}$ according
to the cut-off dependence of $\sigma_{\rm inel}$. 
Fig.~\ref{fig3} illustrates that with this rescaling, the proposed MC algorithm satisfies this important 
property of cut-off independence. 
More explicitly, by changing the values of the IR and/or UV cut-off, 
we change the numerical value of $\sigma_{\rm inel}$ so that  $\lambda_{\rm inel}$ varies like
$\lambda_{\rm inel} \propto 1/\log \left( \omega_{\rm max}/\omega_{\rm min}\right)$. 
Once an inelastic scattering
center is identified in the MC simulation, the kinematics of the emitted gluon is then chosen in the same
kinematic range $\omega \in \left[ \omega_{\rm min} ; \omega_{\rm max}\right]$ that was used for the calculation of
$\sigma_{\rm inel}$. As seen on the right hand side of Fig.~\ref{fig3}, this procedure results in {\it cut-off
independence} of physical results: choosing $\omega_{\rm min}$ and $\omega_{\rm max}$ specifies the range
within which results are generated, but it does not affect the results within this range.
%
\begin{figure}[t]
\begin{center}
\includegraphics[height=10.0cm, angle=0]{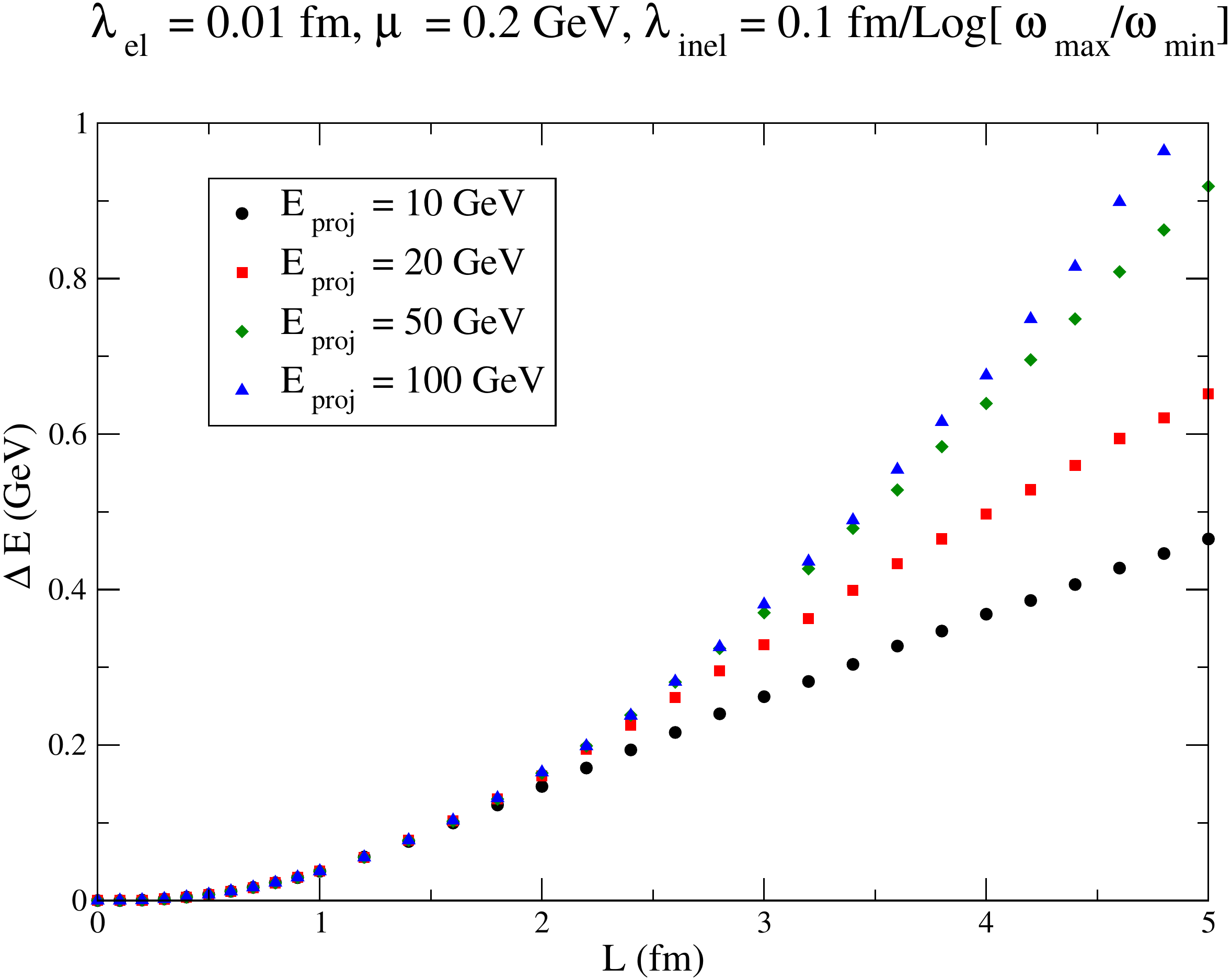}
\caption{
The average energy loss $\Delta E$ as a function of in-medium path length $L$ and for different values of
the UV regulator $\omega_{\rm max}=E_{\rm proj}$ of the differential inelastic cross section. 
}
\label{fig4}
\end{center}
\end{figure}

In general, the appearance of IR and UV cut-offs in the MC algorithm can have different reasons.
For differential inelastic cross sections that implement exact energy-momentum conservation, there is no
need to specify by hand an UV cut-off $\omega_{\rm max}$. Rather, the form of the cross section
will automatically account for the physical requirement that gluons can only be emitted 
with energies smaller than the energy of the incoming partonic projectile, 
$\omega_{\rm max} = E_{\rm proj}$.
The introduction of an UV cut-off is only necessary, since one uses typically the approximate high-energy limit of
the radiation cross section $\propto 1/\omega$, that extends to arbitrarily large gluon energy.
For the case of the IR cut-off $\omega_{\rm min}$ of the $\omega$-integration, or for  the case
of the corresponding IR regulator $\epsilon$ of the ${\bf k}$-differential cross section that enters the 
total inelastic cross section(\ref{eq4.7}) via the factor $f_\text{prop}$, the situation is different. There is no 
perturbative physics argument that could specify the precise value of these regulators. All one can
require is that whatever values for these IR cuts are chosen, the physics simulated above these
values does not depend on the choice of the regulator. Fig.~\ref{fig3} illustrates that this
requirement is satisfied by the proposed MC algorithm. 

\subsection{MC results for the average parton energy loss $\Delta E$}
\label{sec5.3}

In this subsection, we discuss MC simulation results for the average parton energy loss $\Delta E$. The main
purpose of this discussion is to give numerical support to equation (\ref{eq5.8}). 
%
\begin{figure}[t]
\begin{center}
\includegraphics[height=10.0cm, angle=0]{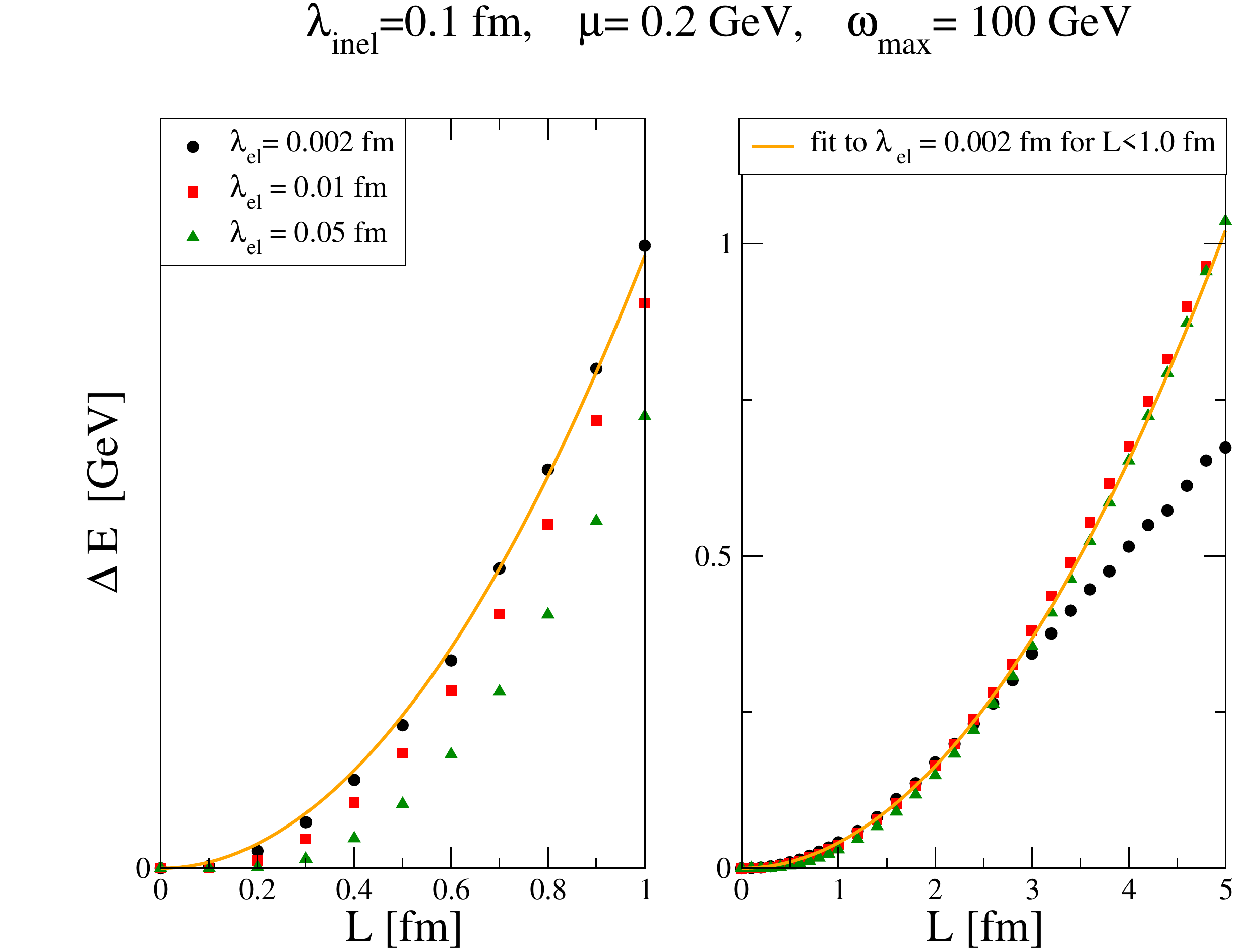}
\caption{
The average parton energy loss $\Delta E(L)$ for different values of the elastic mean free path.
}
\label{fig5}
\end{center}
\end{figure}

Fig.~\ref{fig4} shows the $L$-dependence of the average energy loss for different values of the 
UV regulator $\omega_{\rm max} = E_{\rm proj}$. In this and all subsequent simulations, 
the  value of the inelastic mean free path was adjusted to the varying phase space,
$\lambda_{\rm inel} \propto 1/\log \left( \omega_{\rm max}/\omega_{\rm min}\right)$, 
so that cut-off independent results were obtained.
Since gluons of larger energy $\omega$ require on average a longer in-medium path length to form,
one expects on general grounds that the small-$L$ behavior of the average parton energy loss $\Delta E$ is 
independent of the choice of the UV regulator $\omega_{\rm max}$. 
This is seen clearly in Fig.~\ref{fig4} for sufficiently small $L$. Moreover, for sufficiently
large $L > L_c$, Fig.~\ref{fig4} confirms the expected linear $L$-dependence 
of $\Delta E$. The transition from a quadratic
to a linear dependence occurs at an in-medium path length of order $L_c \propto \sqrt{\omega_{\rm max}}$ that increase with the UV cut-off $\omega_{\rm max} = E_{\rm proj}$. 
Our remark about the accuracy of scale estimates, made about $\omega_c$
in the discussion of Fig.~\ref{fig2}, and in footnote~\ref{foot1}, applies here too. We note in particular that
the quantity $L_c$ is not a quantitative prediction of the BDMPS-Z formalism, but that it characterizes only the
expected parametric dependencies of (\ref{eq2.2}). Consistent with this, we observe that the transition 
from quadratic to linear behavior shows the parametric dependencies expected for  the BDMPS-Z formalism. 
%
\begin{figure}[t]
\begin{center}
\includegraphics[height=10.0cm, angle=0]{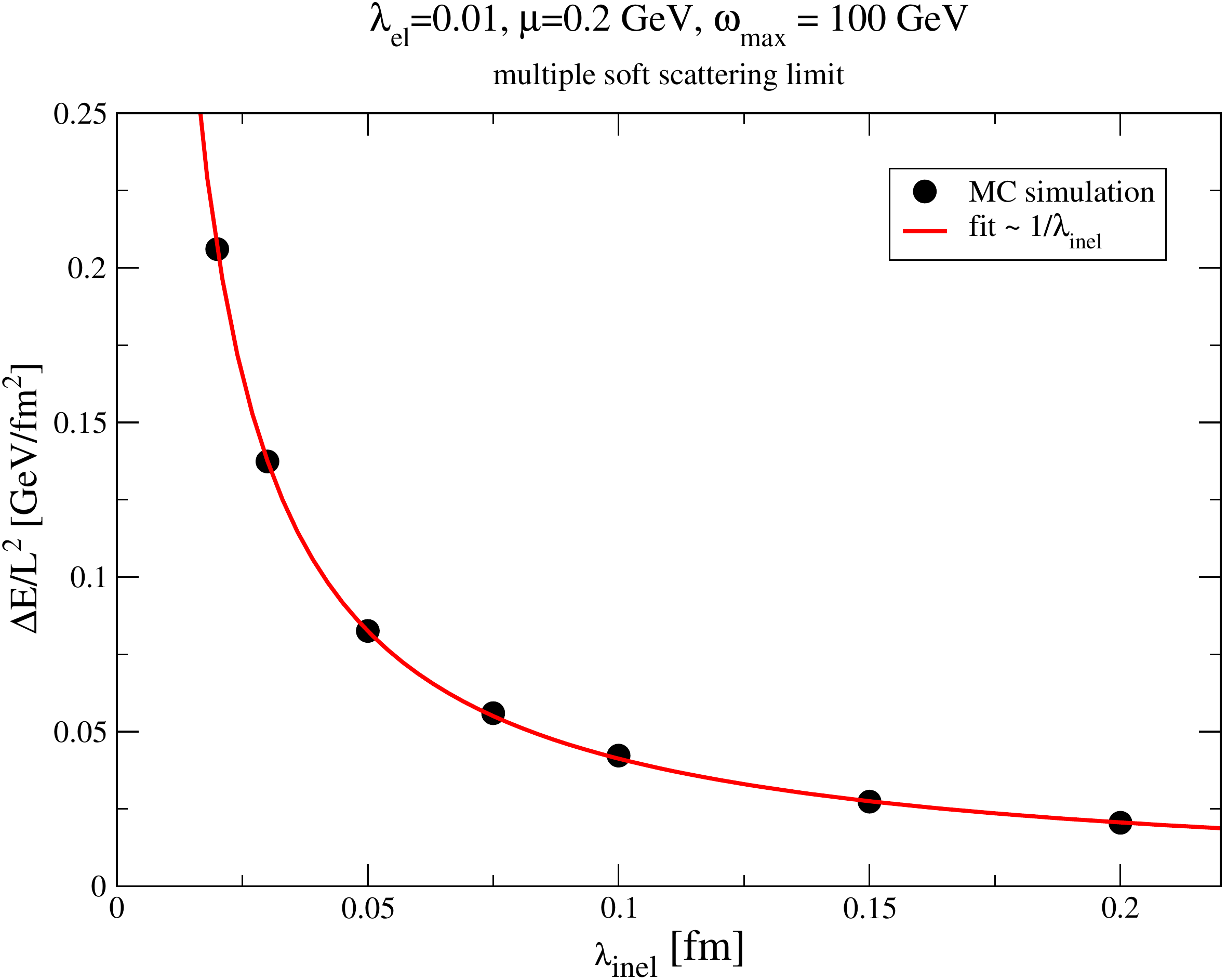}
\caption{
The dependence of the average parton energy loss $\Delta E$ on the inelastic mean free path $\lambda_{\rm inel}$.
}
\label{fig6}
\end{center}
\end{figure}

In general, we find that the ansatz $\Delta E(L) = a\, L^2$ provides a very good description of results of
the MC simulations, if the prefactor $a$ is fit in the range $L < L_c$. But for values $L \ll L_c$, results for 
$\Delta E(L)$ tend to lie significantly below the $L^2$-fit. As we discuss now,
this deviation can be understood by studying the dependence of average parton energy loss on the 
elastic mean free path $\lambda_{\rm el}$, see Fig.~\ref{fig5}. According to equation
(\ref{eq5.8}), $\Delta E \propto \lambda_{\rm el}\, q_{\rm eff}$. Since $q_{\rm eff} = \hat{q} \approx 
\frac{\mu^2}{\lambda_{\rm el}}$, one expects that the average parton energy loss is independent of
$\lambda_{\rm el}$ for $L < L_c$ and for fixed average momentum transfer $\mu$ per scattering center. 
On the other hand, the critical length depends on $\lambda_{\rm el}$, 
$L_c \propto 1/\sqrt{\hat{q}} \propto \sqrt{\lambda_{\rm el}}$
and therefore the $L^2$-dependence of $\Delta E$ should extend to larger values of $L$ for 
larger values of $\lambda_{\rm el}$. On the scale of sufficiently large $L$, these features are 
confirmed by the MC data, see the right hand side of Fig.~\ref{fig5}:  
results fall on a common $L^2$-curve for $L<L_c$, and they turn to a linear $L$-dependence
at a scale $L_c \propto \sqrt{\lambda_{\rm el}}$. 
(The curve for $\lambda_{\rm el} = 0.01$ fm in Fig.~\ref{fig5}
turns to a linear $L$-dependence around $L\sim 5$ fm, while the curve for $\lambda_{\rm el} = 0.05$ fm
shows a quadratic behavior to much larger $L$, data not shown.)

On scales of very small in-medium pathlength, however, there is a characteristic deviation from the 
$\lambda_{\rm el}$-independence of $\Delta E$. On the left hand side of Fig.~\ref{fig5}, we fit an 
$L^2$-dependence to the data obtained for the smallest
elastic mean free path $\lambda_{\rm el} = 0.002$ fm. Remarkably, at large $L$, this fit reproduces perfectly the data
simulated with a 25 times larger mean free path, although this parameter set lies significantly below the $L^2$-curve
for $L<1.0$ fm. This illustrates that increasing $\lambda_{\rm el}$ at fixed $L$ amounts to studying deviations 
from the soft multiple scattering limit. These occur when the probability for no scattering becomes
sizeable -- this is an effect that does not occur in the multiple soft scattering calculation but that 
will always be present in
the Monte Carlo implementation. However, at fixed small value $L$, the characteristic 
$L^2$-dependence of the soft multiple scattering limit (\ref{eq5.8}) can always be recovered by going to sufficiently
small values of $\lambda_{\rm el}$. The default parameter choice $\lambda_{\rm el} = 0.01$ fm used in 
this section was largely motivated by the idea to go sufficiently deep into the multiple scattering
limit $\lambda_{\rm el} \ll \lambda_{\rm inel}$ to observe a quadratic $L$-dependence on a scale
of 1 fm. 
In summary, Fig.~\ref{fig5} confirms the $\lambda_{\rm el}$-dependence of the 
expression (\ref{eq5.8}) for the average parton energy loss and it quantifies the relevance of the multiple soft scattering approximation (\ref{eq5.3}). 

Motivated by these observations, we perform all fits of the quadratic $L$-dependence
of $\Delta E(L)$ in the range $L \in \left[0; L_c \right]$. We can then confirm the other parametric 
dependencies of equation (\ref{eq5.8}). In particular, we have calculated the average parton energy 
loss for different values of the inelastic mean free path $\lambda_{\rm inel}$, 
and we have fit the prefactor $a$ of $\Delta E(L) = a\, L^2$, see Fig.~\ref{fig6}. The average energy loss 
is found to be inversely proportional to $\lambda_{\rm inel}$. 
%
\begin{figure}[h]
\begin{center}
\vskip .5cm
\includegraphics[height=10.0cm, angle=0]{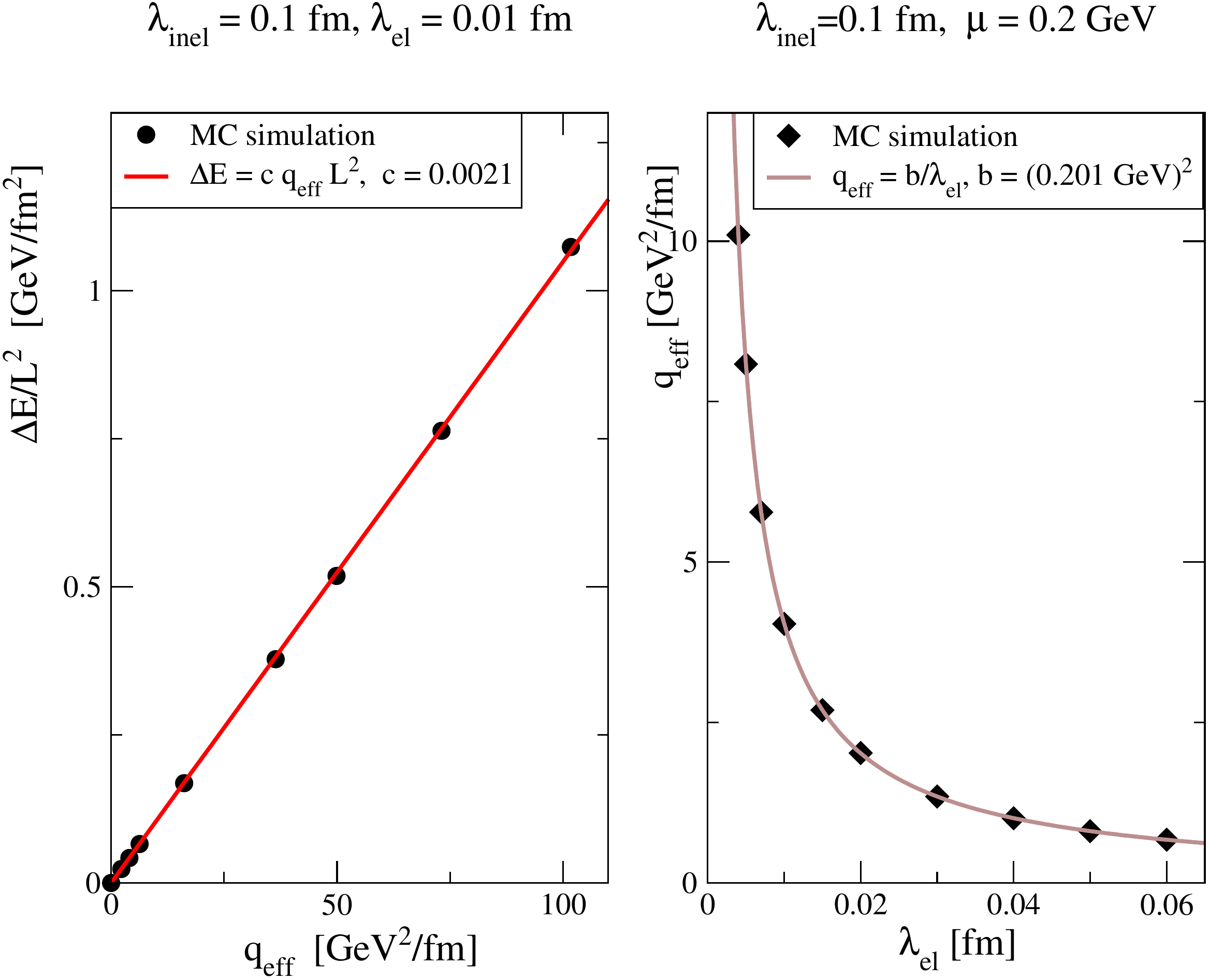}
\caption{
LHS: Dependence of the average parton energy loss on the effective quenching parameter $q_{\rm eff}$.
RHS: The effective quenching parameter as a function of the elastic mean free path $\lambda_{\rm el}$. 
}
\label{fig7}
\end{center}
\end{figure}

We have also characterized the dependence of the average parton energy loss on the quenching parameter.
The right hand side of
Fig.~\ref{fig7} confirms that for the current choice of soft elastic scatterings (\ref{eq5.4}), the effective quenching 
parameter $q_{\rm eff}$ satisfies indeed $q_{\rm eff} = \hat{q} \approx \mu^2/\lambda_{\rm el}$. The 
left hand side of Fig.~\ref{fig7} provides the check that the average parton energy loss depends linearly on 
$q_{\rm eff}$. 

With the figures ~\ref{fig4}, ~\ref{fig5}, ~\ref{fig6} and \ref{fig7}, we have confirmed all parametric dependencies
of the BDMPS-Z result (\ref{eq5.8}) for the average parton energy loss. To determine the overall normalization, 
we make the ansatz $\Delta E = c\, q_{\rm eff}\, L^2$. A direct fit
of $\Delta E = c\, q_{\rm eff}\, L^2$ to MC data in Fig.~\ref{fig7} results in $c= 0.0021$.
On the other hand, we require from (\ref{eq5.8})
\begin{equation}
 c = \frac{1}{4}\, \frac{1}{f_{\rm prop}\, \log\left( \omega_{\rm max} / \omega_{\rm min}\right)} 
	 \frac{\lambda_{\rm el}}{\lambda_{\rm inel}} \equiv 2.1\, 10^{-3}\, .
	 \label{eq5.9}
\end{equation} 
For the values $\lambda_{\rm el}/\lambda_{\rm inel} = 1/10$, and $\log \left(\omega_{\rm max} / \omega_{\rm min}\right)
= \log(100/0.05) \approx 7.6$ used in the simulations of Fig.~\ref{fig7}, we find therefore 
\begin{equation}
	f_{\rm prop} = 1.58\, .
	\label{eq5.10}
\end{equation}
We recall that in the present formulation, the value of $f_{\rm prop}$ is not a prediction of the BDMPS-Z formalism.
Rather, as argued in the discussion of (\ref{eq4.7}), this factor absorbs the remaining dependence on the IR 
cut-off of the total inelastic cross section that is needed in intermediate steps of the MC algorithm.  
It is a prediction, however, that the factor  $f_{\rm prop}$ is of order unity, and that it is a
universal factor that is valid for all model parameter choices. This later statement will be further supported
by the numerical studies in section~\ref{sec5.4}. 

We note that the factor $f_{\rm prop}$ absorbs also uncertainties of the MC implementation. In particular,
we know from further numerical studies that $f_{\rm prop}$  grows roughly proportional 
with $\varphi(\tau_f)$ (data not shown).  Since the choice of $\varphi(\tau_f) = 3$ adopted
here is uncertain by ca. 15{\%} (see discussion of eq. (\ref{eq4.3})), the factor
$f_{\rm prop}$ will also absorb this uncertainty. Noting that in the proposed MC algorithm, 
the acceptance criterion for produced gluons (step 5 in section ~\ref{sec4.2}) is based solely on
the parametric arguments of section~\ref{sec2.3.3}, we have also investigated modifications of this
acceptance criterion. In one extreme alternative version, we required instead that gluons are counted
towards the medium-induced spectrum if their formation after the last momentum 
transfer is completed within a time of scale $L$,
irrespective of whether this amounts to completed formation inside or outside of the medium. 
For this modified MC algorithm, we repeated the entire study of sections~\ref{sec5}
and ~\ref{sec6} with analogous conclusions and very similar figures. The main difference compared 
to the results presented here was that we found an $f_{\rm prop}$ that was approximately a factor 2 
smaller than the value quoted in (\ref{eq5.10}). From this we conclude that depending on how one
implements those elements of the MC algorithm for which the opacity expansion of (\ref{eq2.2})
provides only qualitative but not quantitative guidance, one arrives at a different factor $f_{\rm prop}$ 
of order unity. Most importantly, however, once these ambiguities in the MC implementation are 
fixed by choosing a specific value for $f_{\rm prop}$, the absolute normalization of the simulated parton 
energy loss is fixed for all parameter choices.

\subsection{MC results for phenomenologically motivated parameter values}
\label{sec5.4}
 The choice of elastic and inelastic mean free paths amounts to specifying the strong coupling constant 
 $\alpha_s$, see (\ref{eq5.5}). In the numerical studies in subsections~\ref{sec5.2} and 
 \ref{sec5.3}, we focussed on the perturbative limit $\lambda_{\rm el} \ll \lambda_{\rm inel}$.
 The parameters chosen in these studies correspond to a nominally perturbative regime
 in which $\alpha_s \sim O(10^{-2}- 10^{-3})$ or smaller. 
 We now establish that the properties of the MC 
 algorithm discussed in sections ~\ref{sec5.2} and \ref{sec5.3}, persist for phenomenologically 
 more relevant parameter choices. 
%
\begin{figure}[h]
\begin{center}
\vskip 1.5cm
\includegraphics[height=10.0cm, angle=0]{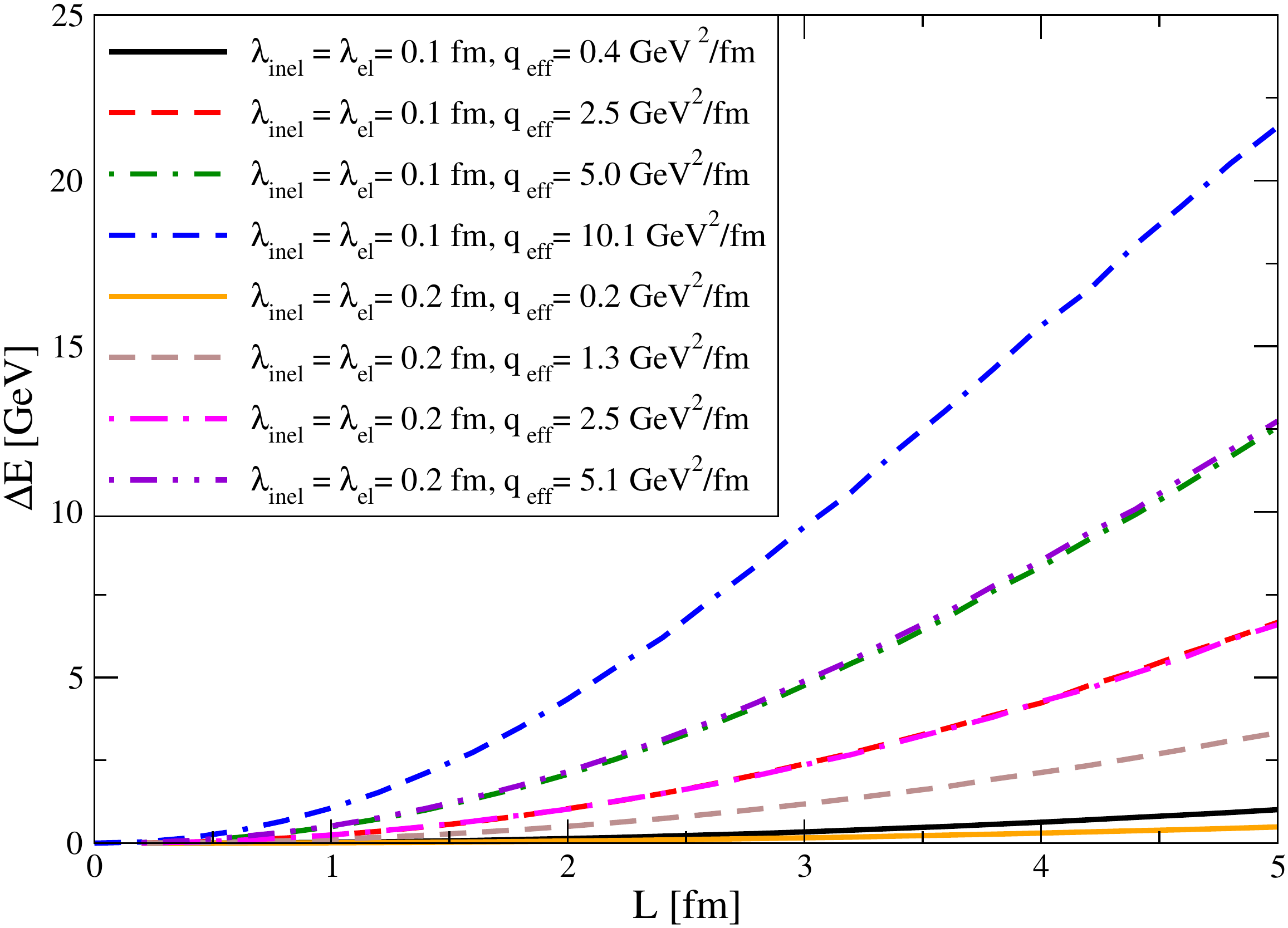}
\caption{
The average medium-induced energy loss $\Delta E$ as a function of in-medium path length $L$
for a quark of energy $100$ GeV,
calculated for choices of $\lambda_{\rm el} = \lambda_{\rm inel}$.}
\label{fig8}
\end{center}
\end{figure}
 
\begin{figure}[h]
\begin{center}
\vskip .5cm
\includegraphics[height=10.0cm, angle=0]{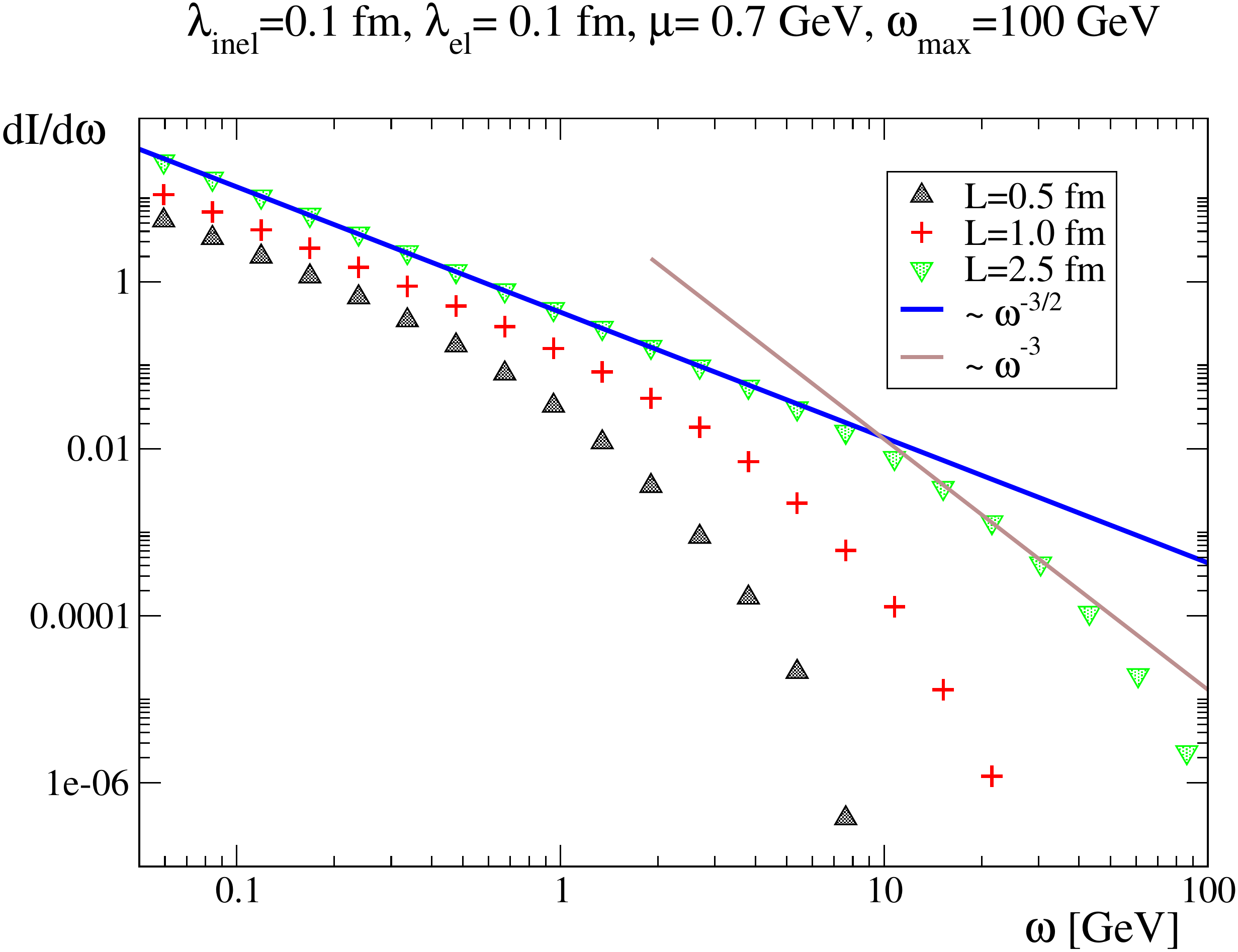}
\caption{
Same as Fig.~\protect\ref{fig2} but for a phenomenologically relevant set of model parameters.}
\label{fig9}
\end{center}
\end{figure}

 According to equation (\ref{eq5.5}), realistic values for $\alpha_s$ are obtained for choices 
 $\lambda_{\rm el} = O(\lambda_{\rm inel})$, and this motivates the parameter choices of the
 simulations shown in Fig.~\ref{fig8} and ~\ref{fig9}. These simulations included gluon 
radiation in the range $\omega_{\rm min} = 50$ MeV to $\omega_{\rm max} = 100$ GeV. We note
that physical results do not depend on the precise choice of $\omega_{\rm min}$; in particular, 
a larger value of $\omega_{\rm min}$ could be absorbed in a rescaled inelastic mean free path
$\lambda_{\rm inel} \propto 1/ \log\left(\omega_{\rm max},\omega_{\rm min}\right)$, as discussed in
the context of Fig.~\ref{fig3}. On the other hand, physical results depend on the upper
boundary $\omega_{\rm max}$ that sets the critical length  $L_c \simeq \sqrt{4\omega_{\rm max}/\hat{q}}$ 
 at which $\Delta E(L)$ changes from a quadratic to a linear $L$-dependence.  The scale
 of $\omega_{\rm max}$ is set by the physical UV cut-off on the radiation spectrum that is given
 by the energy of the partonic projectile. 

For the simulations shown in Fig.~\ref{fig8}, we studied two different values of $\lambda_{\rm inel}
= \lambda_{\rm el}$ for Yukawa masses $\mu = $ $0.2$, $0.5$, $0.7$ and $1.0$ GeV in the 
elastic scattering cross section (\ref{eq4.5}), respectively.  These Yukawa masses set the
scale of the transport coefficient  $q_{\rm eff} \sim \mu^2/\lambda_{\rm el}$.  With these parameter choices, 
medium-induced gluon radiation is studied for a projectile parton of $E_{\rm proj} = 100$ GeV
energy,  propagating through a time-independent static medium of transport coefficient
$q_{\rm eff}$. The results in Fig.~\ref{fig8} shows that the transition from a quadratic to a linear behavior 
occurs also for phenomenologically relevant parameter values at a scale of order $L_c$, as established 
in section~\ref{sec5.2} in the multiple soft scattering limit. Fitting a quadratic dependence to the 
small-$L$ region of $\Delta E(L)$, we confirm all parametric dependencies of (\ref{eq5.8}).
Moreover, we confirm within an accuracy of better than 5 \%, that the proportionality factor $f_{\rm prop}$ 
of (\ref{eq5.8}) takes the same value as in the multiple soft scattering limit, $f_{\rm prop} = 1.58$. 
This shows that for one universal normalization of the inelastic cross section (\ref{eq4.7}),
the MC algorithm accounts faithfully for the results of the BDMPS-Z formalism (\ref{eq2.2}) over a very 
wide parameter range, including phenomenologically motivated parameter choices. 
%

The BDMPS-Z path integral (\ref{eq2.2}) does not depend separately on the coupling constant,
the number of scattering centers per unit path length $n$ and the dipole cross section $\sigma$. Rather,
it depends only on $\alpha_s$ and on the linear combination $n\, \sigma$. As a consequence, the MC 
implementation of (\ref{eq2.2}) does not depend separately on $q_{\rm eff}$, $\lambda_{\rm el}$ and
$\lambda_{\rm inel}$. Rather, it depends only on two combinations of these three parameters, 
which may be chosen to be $q_{\rm eff}$ and $\lambda_{\rm el}/\lambda_{\rm inel}$, only. 
This is clearly seen in Fig.~\ref{fig8}, where the choices $\lambda_{\rm el} = \lambda_{\rm inel} = 0.1$ fm
with $\mu = 0.7$ GeV  and $\lambda_{\rm el} = \lambda_{\rm inel} = 0.2$ fm with $\mu = 1.0$ GeV 
correspond to different microscopic pictures of the interaction between projectile and medium,
but result both in the same average squared momentum transfer per unit path
length $q_{\rm eff} = 5$ GeV$^2$/fm, and in the same average parton energy 
loss.  

Comparing Fig.~\ref{fig9} to Fig.~\ref{fig2}, we observe that also the $\omega$-differential information
continues to show for phenomenologically motivated parameter choices the main features 
that we have established in the multiple soft scattering limit. In particular, the spectrum shows for soft
gluon energies the $\omega^{-3/2}$-dependence characteristic for medium-induced coherence,
and for large gluon energy a steeper fall-off $\propto 1/\omega^3$. Also, the transition between 
these two limiting spectra occurs at the scale $\omega \sim \omega_c = \frac{1}{2} \hat{q}\, L^2$, as expected from
the BDMPS-Z result (\ref{eq5.1}).

We finally note that for the model parameters chosen in this subsection, one finds the still rather small
value of $\alpha_s \approx 0.1$. For larger values of $\alpha_s$, the resulting average parton energy loss
will increase correspondingly. Therefore, Fig.~\ref{fig8} illustrates that for phenomenologically relevant
parameters and length scales, the average parton energy loss can attain values of tens of GeV.

\section{Numerical results on transverse momentum broadening}
\label{sec6}
In section~\ref{sec5}, we have demonstrated that the MC algorithm of subsection~\ref{sec4.1} reproduces 
faithfully the $\omega$-dependence of (\ref{eq2.2}) for ${\bf k}$-integrated quantities.
We now discuss how the MC algorithm accounts for the ${\bf k}$-dependence of the 
BDMPS-Z formalism. 

%
\begin{figure}[h]
\begin{center}
\vskip .5cm
\includegraphics[height=10.0cm, angle=0]{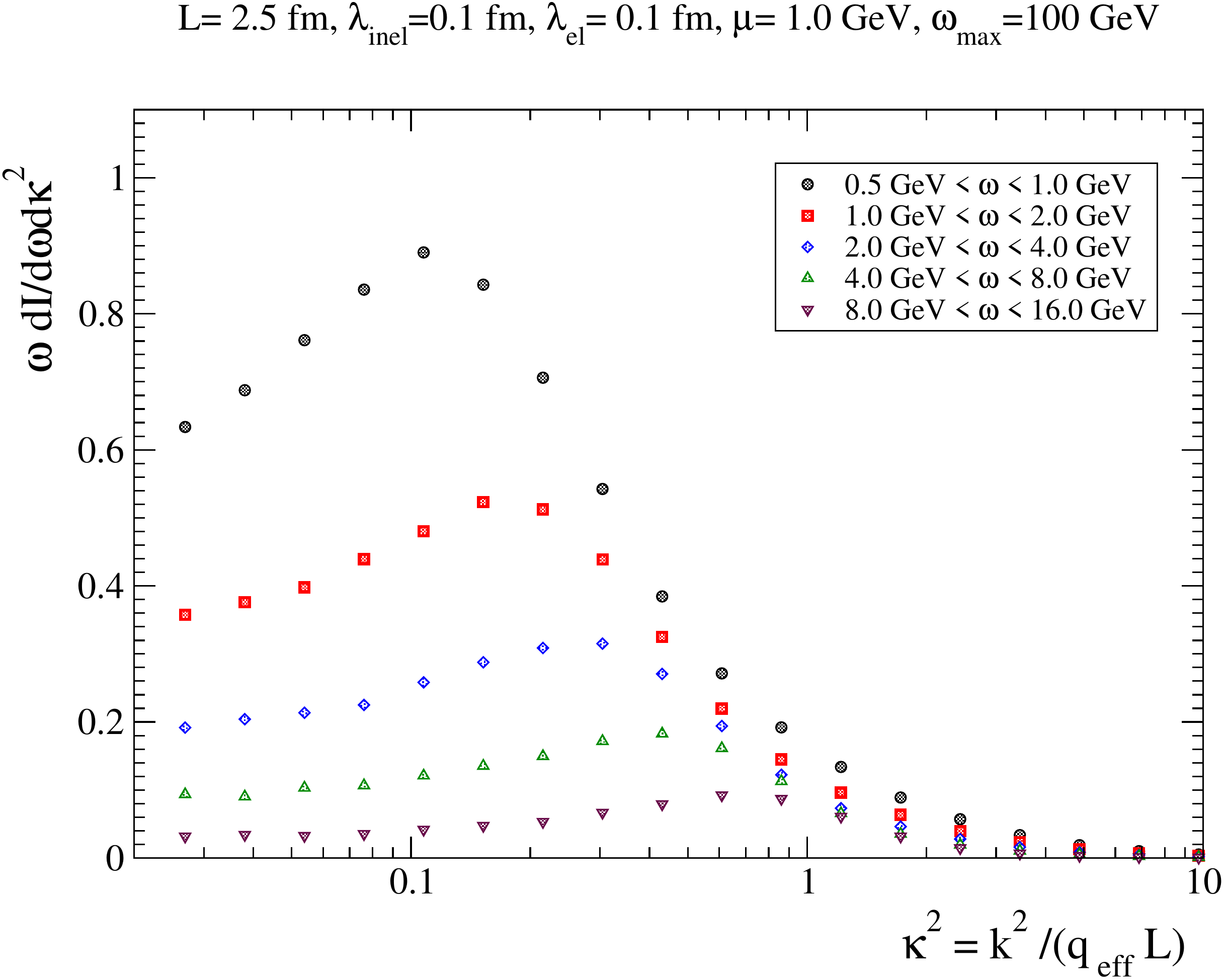}
\caption{The distribution of medium-induced gluons as a function of the normalized
squared transverse momentum $\kappa^2 = {\bf k}^2/ q_{\rm eff} L$. Data points display the simulated 
gluon yield separately for different ranges of gluon energy $\omega$.
}
\label{fig10}
\end{center}
\end{figure}

It is a generic feature of the BDMPS-Z formalism that the transverse momentum of produced
gluons is accumulated according to transverse Brownian motion, 
\begin{equation}
	\langle {\bf k}^2\rangle \propto \hat q\, L\, .
	\label{eq6.1}
\end{equation}
To identify this feature numerically, we plot in Fig.~\ref{fig10} the simulated double differential distribution 
$\frac{\dr I}{\dr\omega\, \dr{\bf k}}$ for different ranges of gluon energies
$\omega$ as a function of $\kappa^2 = {\bf k}^2/ q_{\rm eff} L$. In accordance with (\ref{eq6.1}), the
main contribution to the yields of simulated gluons lies in the range $\kappa^2 \leq1$, irrespective of
the gluon energy, and irrespective of the choice of the parameters $\lambda_{\rm el}$, $\lambda_{\rm inel}$
and $\mu^2$ that control the rate of gluon production and its transverse momentum broadening. 
The double differential distribution $\omega \frac{\dr I}{\dr \omega\, \dr\kappa^2}$ 
of (\ref{eq2.2}) has been analyzed and plotted for the soft multiple scattering limit and the $N=1$
opacity approximation in Ref.~\cite{Salgado:2003rv}. We note 
that the results of the MC simulation shown in Fig.~\ref{fig10} reproduce very well the main
results of Ref.~\cite{Salgado:2003rv}. In particular, the gluon yield dies out at a scale $\kappa^2 \sim 
O(1)$, it decreases with increasing gluon energy, and it shows a plateau for logarithmically small values
of $\kappa^2$. Also, the overall normalization of the MC results for the double differential distribution 
is in general agreement with the results of Ref.~\cite{Salgado:2003rv}. 

%
\begin{figure}[h]
\begin{center}
\vskip .5cm
\includegraphics[height=12.0cm, angle=0]{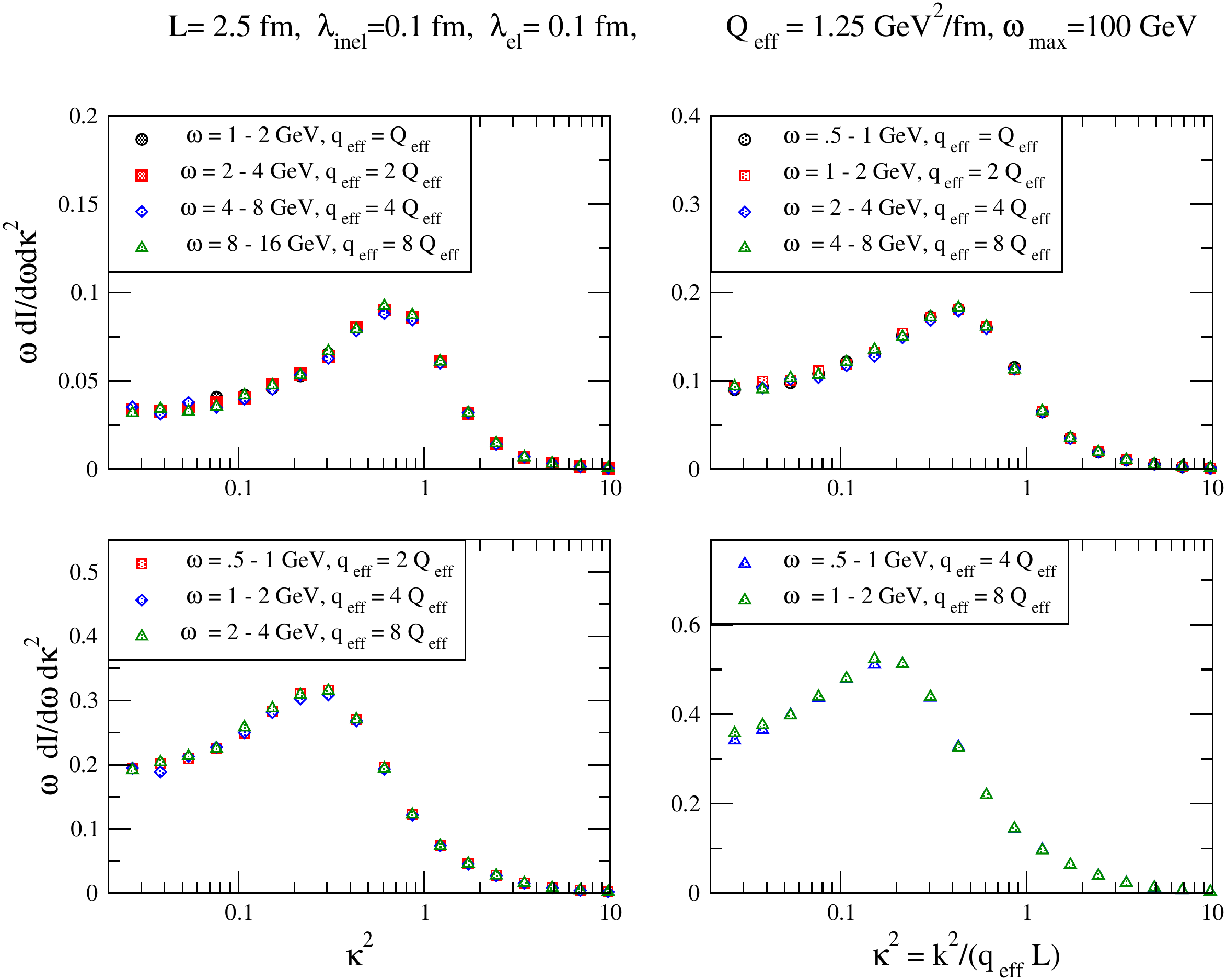}
\caption{Same as Fig.~\protect\ref{fig10}, but for different ranges of gluon energy 
$\omega \in \left[ \omega_{\rm min}; \omega_{\rm max} \right]$, and for different
values of the quenching parameter $q_{\rm eff}$. That MC results for different
parameter choices fall on a universal curve illustrates the scaling property 
(\protect\ref{eq6.2}). }
\label{fig11}
\end{center}
\end{figure}

There are also qualitatively noteworthy though quantitatively small differences between 
the analysis of (\ref{eq2.2}) in Ref.~\cite{Salgado:2003rv} and the output of the
MC algorithm proposed here. In particular, destructive 
medium-induced interference effects can modify the gluon radiation such that in comparison to the 
vacuum distribution, the total yield of produced gluons is reduced in some phase space region
below its average value in the vacuum. This would show up in negative values of the 
medium-induced gluon energy distribution $\omega \frac{dI}{d\omega\, d{\bf k}}$. 
Such an effect has been identified 
indeed in a small phase space region of (\ref{eq2.2})~\cite{Salgado:2003rv}. A similar observation
of small negative contributions has been made for the ${\bf k}$-integrated distribution $\omega
\frac{\dr I}{\dr\omega}$ at very small in-medium path length. 
In contrast to the analytic calculation, which considers both vacuum and medium induced radiation and
subtracts the unperturbed vacuum spectrum from the total gluon spectrum, the MC algorithm neglects the vacuum emissions
(as a consequence the MC spectrum cannot become negative).
While this
can be seen in small deviations of (\ref{eq2.2}) from MC results, we emphasize here that 
all numerically important, generic features of (\ref{eq2.2}) are accounted for quantitatively by the
MC algorithm. 
%
\begin{figure}[h]
\begin{center}
\vskip .5cm
\includegraphics[height=10.0cm, angle=0]{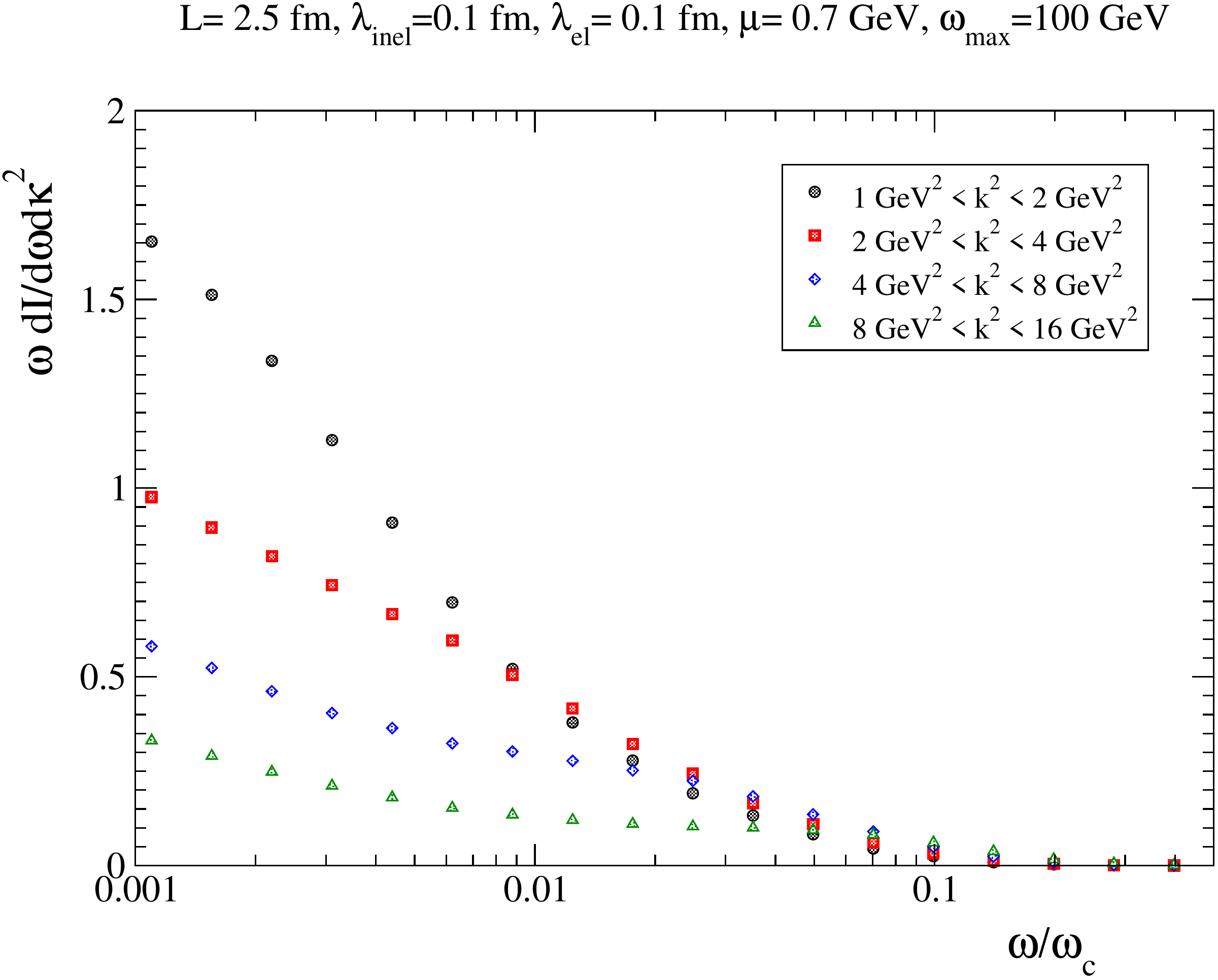}
\caption{The distribution of medium-induced gluons as a function of gluon energy $\omega$.
Data points display the simulated gluon yield separately for different ranges of 
transverse gluon momentum  ${\bf k}$.
}
\label{fig12}
\end{center}
\end{figure}

It is also a characteristic feature of the BDMPS-Z formalism that medium-induced gluon radiation
occurs for all gluon energies that accumulate at least a phase factor of order unity in the medium,
$\frac{\langle {\bf k}^2\rangle L}{2\, \omega} = \frac{\omega_c}{\omega} > 1$. In combination with 
the transverse momentum broadening (\ref{eq6.1}), the technical manifestation of this statement
is that the gluon radiation spectrum is only a function of the rescaled variables $\omega/\omega_c$ and
${\bf k}^2/ \hat{q} L$,
\begin{equation}
 \omega \frac{\dr I}{\dr\omega\, \dr{\kappa^2}} = f\left( \frac{\omega_c}{\omega},
    \frac{{\bf k}^2}{ \hat{q} L} \right)\, .
    \label{eq6.2}
\end{equation}
To illustrate that this scaling is satisfied by the proposed MC algorithm, we have plotted in Fig.~\ref{fig11}
MC simulations of $\omega \frac{\dr I}{\dr \omega\, \dr{\kappa^2}} $ as a function of 
$\kappa^2$ for different ranges of gluon energy, $\omega \in \left[\omega_{\rm min};\omega_{\rm max} \right]$,
and for different values of $\mu^2$ (i.e. different values of $q_{\rm eff}$), keeping $L$, $\lambda_{\rm el}$
and $\lambda_{\rm inel}$ fixed. It is then a direct consequence of (\ref{eq6.2}) that 
varying $\omega_{\rm min}$, $\omega_{\rm max}$ and $q_{\rm eff}$ by the same factor will leave
the distribution $\omega \frac{\dr I}{\dr\omega\, \dr{\kappa^2}} $ unchanged. Fig. ~\ref{fig11} illustrates
that this generic scaling property of the BDMPS-Z formalism is satisfied by the MC algorithm. 

We finally discuss the $\omega$-dependence of the double-differential gluon energy distribution 
for fixed values of $\kappa^2$. The multiple soft scattering approximation and the $N=1$ opacity
approximation of (\ref{eq2.2}) are known to result in an $\omega$-dependence of 
$\omega \frac{\dr I}{\dr\omega\, \dr{\kappa^2}}$ that is flatter for  increasing $\kappa^2$~\cite{Salgado:2003rv}.
The same feature is clearly seen in Fig.~\ref{fig12}.

\begin{figure}[h]
\begin{center}
\vskip .5cm
\includegraphics[height=12.0cm, angle=0]{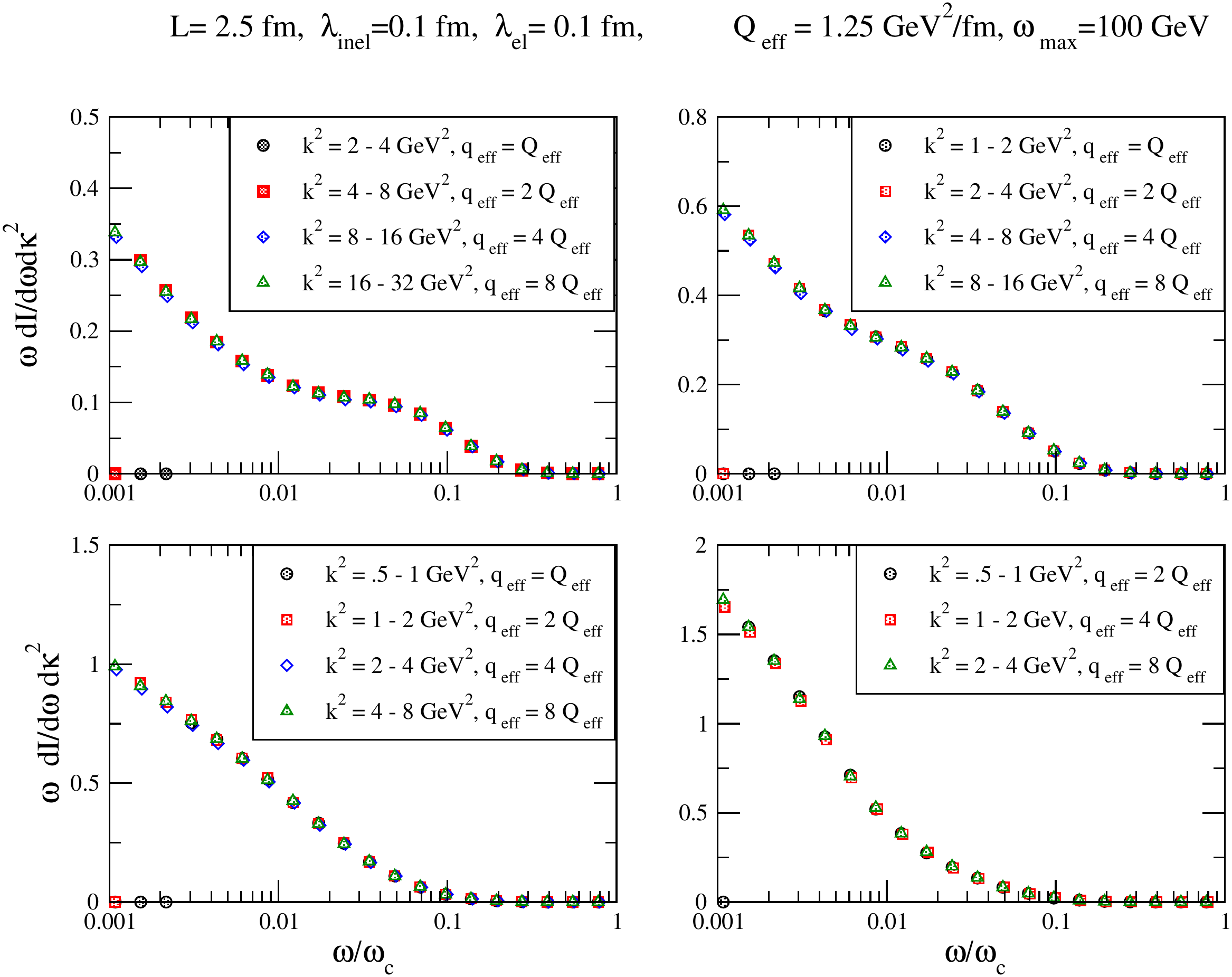}
\caption{Same as Fig.~\protect\ref{fig11}, but for different ranges of squared transverse momentum
${\bf k}^2$, and for different
values of the quenching parameter $q_{\rm eff}$. That MC results for different
parameter choices fall on a universal curve illustrates the scaling property (\protect\ref{eq6.2}).
}
\label{fig13}
\end{center}
\end{figure}

We finally show in Fig.~\ref{fig13} that the universal scaling property (\ref{eq6.2}) is also clearly 
supported by the analysis of the $\omega$-dependence of $\omega \frac{\dr I}{\dr\omega\, \dr{\kappa^2}}$.
In summary, we conclude that the MC algorithm proposed in section~\ref{sec4} 
reproduces all numerically relevant qualitative and quantitative features of the BDMPS-Z
formalism.

%

\section{Conclusions and Outlook}
\label{sec7}
Multi-parton production processes exhibit destructive quantum interference effects.
In general, their probabilistic implementation involves approximations. For multiple
parton branching in the vacuum, the dominant destructive interference effect
can be taken into account probabilistically by an angular ordering prescription.
This probabilistic reformulation of the analytical expression is an approximation 
that is known to have the same parametric accuracy in $\log Q^2$ and $\log 1/x$ 
as the leading order perturbative calculation. 
Within QCD matter, parton splitting close to the eikonal limit (\ref{eq2.1}) is dominated
by medium-induced destructive quantum interference effects that are calculated in the 
BDMPS-Z formalism. In this paper, we have demonstrated that the dominant medium-induced
interference effect for ${\bf k}$- and $\omega$-differential parton distributions can be taken into
account probabilistically by a re-weighting of gluon emission probabilities based on gluon
formation times. This probabilistic formulation is an approximation of the analytical
BDMPS-Z result (\ref{eq2.2}). We have established in a detailed numerical study to what 
extent it is a good approximation.

The proposed probabilistic implementation of the BDMPS-Z formalism is based on 
approximating by theta-functions those oscillatory functions that interpolate in the analytical
BDMPS-Z formalism between the coherent and incoherent limiting cases. The proposed
MC implementation reproduces by construction the known coherent and incoherent limiting
cases and it interpolates, by construction, between these
limits on the correct momentum scales. This ensures that the MC simulations agree
in normalization and parametric dependencies with the analytically known results. 
In small regions of phase space and for very small in-medium path lengths, however,
destructive interference effects are known to show up in the medium-induced part of the 
${\bf k}$-integrated gluon energy distribution eq.~(\ref{eq2.2}) as oscillatory behavior.
The approximations in the probabilistic reformulation will 
not account for detailed interference effects such as oscillations in $\omega \frac{\dr I}{\dr\omega}$, 
but these are known to be numerically small and they depend also in analytical
calculations on the approximations employed to evaluate eq.~(\ref{eq2.2}). While we have not
advanced a parametric argument for the accuracy of the proposed MC algorithm, we conclude
from the detailed numerical study in section~\ref{sec5} that the algorithm 
allows to implement probabilistically and in a quantitatively controlled manner
all numerically relevant features of the BDMPS-Z formalism.
This includes the correct normalization of the average parton energy loss and the norm and
shape of the $\omega$-differential distribution, as well as the parametric dependencies on 
in-medium path length, transport coefficient and coupling constant. 
An analogous remark applies to the ${\bf k}$-differential distribution, as established in section~\ref{sec6}.

The phenomenological modeling of jet quenching based on the BDMPS-Z formalism faces se\-ve\-ral 
longstanding problems. First, phenomenological models must account for medium-induced
gluon splitting also outside the kinematic regions $E \gg \omega$ and $\omega \gg \vert {\bf k}\vert$,
within which the BDMPS-Z formalism has been derived. Second, energy and momentum is not conserved 
in the BDMPS-Z formalism but its conservation at each microscopic interaction is phenomenologically
important, in particular if it comes to simulating not only leading hadrons but the energy loss (a.k.a. 
medium-modified fragmentation) of reconstructed jets. Third, it is desirable to understand
better how the medium-induced gluon radiation depends on properties of the scattering centers in
the medium, and this requires the ability to vary the nature of the scattering centers in model calculations.
Fourth, it is indispensable for a phenomenological model of medium-modified jet fragmentation
that all components of the parton shower are treated on the same footing,  and that means that all
components can be subject to both elastic and inelastic interactions. In the BDMPS-Z formalism,
however, radiated gluons scatter only elastically, and the projectile quark scatters only inelastically.
Therefore the distribution of subleading partons obtained from the BDMPS-Z formalism must not
be regarded as a suitable proxy for a medium-modified parton shower. Fifth, since the BDMPS-Z
formalism has been derived close to the eikonal approximation, it is recoilness. This has resulted in
a debate that distinguishes in an ad hoc way between collisional and radiative parton energy loss,
rather than pushing for a physical formulation of the problem in which radiative contributions are
necessarily accompanied by recoil (and therefore by effects that one typically associates with
collisional energy loss). 

The possibilities for improving on these major deficiencies of the BDMPS-Z formalism with refined 
analytical techniques appear to be limited. The proposed MC implementation of the 
BDMPS-Z formalism opens significant novel opportunities to this end. In particular, the proposed
algorithm can be supplemented naturally with i) exact kinematics outside the region 
$E \gg \omega \gg \vert {\bf k}\vert$, ii) exact energy-momentum conservation at each interaction
with the medium, iii) a large variety of models for the interaction with between projectile and medium,
iv) a democratic treatment of all components of the parton shower and v) a kinematically correct,
dynamical inclusion of recoil effects. In close analogy to the more mature situation in elementary particle
physics, we expect that MC techniques will become in the next years also in heavy ion physics the 
preferred choice for the description of high-$p_T$ multi-particle final states, and we recognize their 
advantages in interfacing dynamical simulations of parton evolution with hadronization models.
In the present paper, we have demonstrated only that the proposed MC algorithm implements 
all numerically relevant features of the BDMPS-Z formalism probabilistically. In our view, the main
importance of this result lies in the fact that it establishes a starting point for going beyond the
BDMPS-Z formalism in a framework that remains rooted in the analytically identified medium-induced
interference effects. We plan to explore this approach in subsequent work.

%

\acknowledgments
We acknowledge financial support from the European Network MCnet
and from the Extreme Matter Institute EMMI at various stages of this  
work.

\appendix
\section{Formation time of vacuum radiation from the BDMPS-Z formalism} 
\label{appa}
Here, we demonstrate that a simple extension of the opacity expansion of section~\ref{sec2} 
allows one to identify within the BDMPS-Z formalism a formation time for vacuum radiation.
Although this quantity does not enter the MC algorithm proposed in the present paper, we find this
observation sufficiently interesting to discuss it in the present appendix.

The main idea of the following is to gain further insight into the different roles of vacuum and medium-induced
radiation by introducing a length scale $\bar{L}$ that separates the production of the partonic 
projectile at $\xi_0=0$ from its in-medium propagation after time $\bar{L}$. To this end, we study 
the BDMPS-Z formalism for a uniform distribution of scattering centers in a spatial region that is 
separated by a length $\bar{L}$ from $\xi_0$, 
\begin{equation}
	n(\xi) =  \Bigg \{ 
	\begin{array}{l}
	n_0\, ,\qquad \hbox{\rm for $\bar{L} < \xi < \bar{L} + L $\, ,} \\  \\
	0\, ,\qquad \hbox{\rm for $\xi < \bar{L}$ or $ \xi>  \bar{L} + L$\, .}
	\end{array} 
	\label{eq2.35}
\end{equation}
From equation (\ref{eq2.2}), we find then to first order in opacity the medium-induced gluon energy distribution 
\begin{eqnarray}
  \omega {\dr I(N=1)\over \dr \omega\, \dr{\bf k}\, \dr{\bf q}}
  &=& {\alpha_s\, C_R\over \pi^2}\, 
  \frac{1}{(2\pi)^2}\,
    \left( |A({\bf q})|^2 - V_{\rm tot}\, \bar \delta({\bf q}) \right)\, 
    \left[  \frac{1}{\left({\bf k}+{\bf q}\right)^2} 
    	+ \frac{{\bf q}^2}{{\bf k}^2\, \left({\bf k}+{\bf q}\right)^2} 
\right]\, ,
    	\nonumber \\
	&& \qquad \qquad \times
    \,  \left(n_0\, L\right)\, \frac{L\, Q_1 - \sin\left((L+\bar{L})\, Q_1 \right) + \sin\left(\bar{L}\, Q_1 \right) }{L\, Q_1}\, .
  \label{eq2.36}
\end{eqnarray} 
Here, ${\bf k}$ denotes the transverse momentum of the gluon in the {\it final} state, and  $\left({\bf k}+{\bf q}\right)$ can be
regarded as the transverse momentum of an incoming gluonic component of the partonic projectile. The value 
$Q_1 = \left({\bf k}+{\bf q}\right)^2/2\omega$ denotes then the transverse energy of this {\it initial}
gluonic projectile component, prior to exchanging a transverse momentum  ${\bf q}$
with the medium. In the following, we investigate under which conditions this initial
gluonic component can be freed (i.e. radiated) 
by a medium positioned between $\bar{L}$ and $\bar{L}+L$. 

We note first that the vacuum radiation term $H\left({\bf k}+{\bf q}\right)$ in the first line of 
 (\ref{eq2.36}) displays the standard collinear singularity of the vacuum radiation. Also, the 
 medium-induced radiation term $R\left({\bf k},{\bf q}\right)$ shows singularities for vanishing incoming 
gluon momentum $\left({\bf k}+{\bf q}\right)$ and for vanishing outgoing gluon momentum
${\bf k}$,  as one expects for the radiation from an isolated single scattering center. 
We now discuss how the destructive interference term in the second line of (\ref{eq2.36}) regulates the 
 incoming singularity at a scale that depends on the position and thickness of the target. 
 We consider first a medium of a fixed number of active scattering centers, that means, a medium
 of fixed opacity ($n_0\, L = {\rm fixed}$). 
 For gluons of initial 
transverse energy $Q_1$, we can then always find a sufficiently large in-medium path length 
$L \gg 1/Q_1$, so that these gluons can be freed with negligible destructive interference effects,
\begin{eqnarray}
 \frac{L\, Q_1 - \sin\left((L+\bar{L})\, Q_1 \right) + \sin\left(\bar{L}\, Q_1 \right) }{L\, Q_1}
 \Bigg\vert_{L\, Q_1 \gg 1} 
 = 1\, .
 \label{eq2.37}
\end{eqnarray}
What happens in the opposite limit, when the longitudinal extension of the medium $L$ is small
compared to the inverse transverse energy of the incoming gluon, $L \ll 1/Q_1$? 
Expanding the phase factor for $ \left(L\, Q_1\right) \ll 1$, we find 
\begin{eqnarray}
 \frac{L\, Q_1 - \sin\left((L+\bar{L})\, Q_1 \right) + \sin\left(\bar{L}\, Q_1 \right) }{L\, Q_1}\
 &=& \left(1-\cos(\bar{L}Q_1)\right) + \frac{1}{2} \sin\bar{L}Q_1 \, \left(L\, Q_1\right)
 	\nonumber \\
&& + \frac{1}{6} \cos (\bar{L}Q_1) \, \left(L\, Q_1\right)^2  + O\left(L^3\, Q_1^3\right) \, .
\label{eq2.38}
\end{eqnarray}
The limit $n_0\, L$ = fixed, $L\to 0$ corresponds to localizing medium effects exactly at a
distance $\bar{L}$ after the starting point $\xi_0$ of the parton evolution. In this limit, the
phase factor (\ref{eq2.38}) is $ \left(1-\cos (\bar{L}Q_1)\right) $, and it cancels the
$1/\left({\bf k}+{\bf q}\right)^2$ divergencies in (\ref{eq2.36}) only if $\bar{L}Q_1 \ll 1$. 
Therefore, gluons with initial transverse energy $Q_1$ can only be produced in interactions
with the medium, if the medium is placed at a distance
\begin{equation}
	\bar{L} > \frac{1}{Q_1}\equiv \tau_f^{\rm (vac)}\, .
	\label{eq2.39}
\end{equation}
We note that the limit $n_0\, L$ = fixed, $L\to 0$ can be viewed as a gedankenexperiment, 
according to which one produces a parton at time $\xi_0$ and allows for its vacuum evolution up to a time
$\bar{L}$ before testing the content of the evolved vacuum wave function by an interaction with 
the medium at time $\bar{L}$. The inequality (\ref{eq2.39}) suggests a probabilistic picture
according to which - irrespective of the nature of the
medium and the strength of its interaction - one can interact with gluons of transverse energy $Q_1$ 
in the incoming vacuum wave function of the projectile only at times later than $1/Q_1$. In this sense, 
the inverse transverse energy $1/Q_1$ of the gluonic components prior to interaction with the medium has a natural
interpretation as the formation time $\tau_f^{\rm (vac)}$ of gluons in the vacuum.  

Heuristic proposals for the life time of a parent parton in the vacuum are often 
based on its virtuality $Q$. In its own rest frame, a state of virtuality $Q$ is expected to have 
a lifetime $\sim 1/Q$.  In a Lorentz frame in which this virtual partonic state has energy $E$,
its life time $\sim 1/Q$ is Lorentz dilated by a boost factor $E/Q$, 
\begin{equation}
	\tau_{\rm life} \sim \frac{E}{Q^2}\, .
\end{equation}
We consider now the standard perturbative situation that the virtual parent parton
splits into two partons with much lower virtuality and with
momentum fractions $z$ and $(1-z)$ respectively. The relative
transverse momentum ${\bf k}_{\rm pair}$ between the two daughter partons satisfies
then ${\bf k}_{\rm pair}^2 \simeq z\, (1-z)\, Q^2$. Taking the softer daughter parton to be the
gluon with energy $\omega = z\, E$ and $(1-z) \simeq 1$, one finds
\begin{equation}
	\tau_{\rm life} \equiv \frac{E}{Q^2} \simeq 
	\frac{\omega}{{\bf k}_{\rm pair}^2} \propto \tau_f^{\rm (vac)}\, .
	\label{eq2.30}
\end{equation}
To sum up: We have advanced heuristic arguments to characterize the partonic lifetime
of the parent parton by  $\sim E/Q^2$. We now find that this estimate is fully equivalent to 
the formation time of the daughter parton, that we have identified within the
BDMPS-Z formalism in (\ref{eq2.39}) in terms of a transverse gluon energy. 
The BDMPS-Z formalism does not provide an
explicit description for the virtuality evolution since it is limited to the calculation of single
medium-induced gluon emissions. However, the BDMPS-Z formalism knows about the 
virtuality of parent partons in the sense that i) it allows for parton splitting in the absence
of medium effects and ii) it attributes time scales to the vacuum splittings that are consistent
with standard heuristic arguments based on the virtuality of parent partons.


\end{document}